\documentclass[preprint,floats,twoside,floatfix]{revtex4-1}
\usepackage{epsfig,amsmath,amssymb,graphicx,color,calc}

\usepackage{graphicx}
\usepackage{amsmath,amssymb}
\usepackage{psfrag}

\def\be{\begin{equation}}
\def\ee{\end{equation}}
\def\bea{\begin{eqnarray}}
\def\eea{\end{eqnarray}}

\begin{document}

\title{Universality Classes of Critical Points in Constrained Glasses
}

 \author{Silvio Franz (1) and Giorgio Parisi (2)}

 \affiliation{(1) Laboratoire de Physique Th\'eorique et Mod\`eles
   Statistiques, \\ CNRS et Universit\'e Paris-Sud 11, UMR8626,
   B\^at. 100, 91405 Orsay Cedex, France
\\
(2) Dipartimento di Fisica, Universit\`a di  Roma La Sapienza,\\ 
INFN, Sezione di Roma I,  IPFC - CNR, P.le Aldo Moro 2, I-00185 Roma, Italy
}
\begin{abstract} 
  We analyze critical points that can be induced in glassy systems by
  the presence of constraints. These critical points are predicted by
  the Mean Field Thermodynamic approach and they are precursors of the
  standard glass transition in absence of constraints. Through a deep
  analysis of the soft modes appearing in the replica field theory we
  can establish the universality class of these points. In the case of
  the ``annealed potential'' of a symmetric coupling between two
  copies of the system, the critical point is in the Ising
  universality class. More interestingly, is the case of the
  ``quenched potential'' where the a single copy is coupled with an
  equilibrium reference configuration, or the ``pinned particle'' case
  where a fraction of particles is frozen in fixed positions. In these
  cases we find the Random Field Ising Model (RFIM) universality
  class. The effective random field is a ``self-generated'' disorder
  that reflects the random choice of the reference configuration. The
  RFIM representation of the critical theory predicts non-trivial
  relations governing the leading singular behavior of relevant
  correlation functions, that can be tested in numerical simulations.
\end{abstract}

\maketitle

\section{Introduction}
Recent times have seen a renewed interest for glassy systems in
presence of constraints. Glassy relaxation in liquids is dominated by
the presence of metastable states. According to the Mean Field picture
of the glass transition \cite{Ca09,WL12}, also known as Random First
Order Transition (RFOT) \cite{KTW89}, these states have a well defined
thermodynamic meaning and can be probed and stabilized by imposing
suitable constraints that modify the Hamiltonian. The simplest
procedure consists in considering two copies of the system and
introduce an attraction between the particles of the first and the
second copy \cite{FPV92}. The free-energy as a function of the
overlap, which is the conjugate parameter to the strength of the
attraction, is often referred to as the ``annealed potential''
function. A second, more refined procedure consists in fixing a
reference configuration and biasing the Boltzmann probability of the
system in the direction of this configuration \cite{FP95}. One can
consider an external potential that provides an attraction for the
particles of the system to the position they take in the reference
configuration. In this case, the free-energy as a function of the
overlap is called ``quenched potential''. Finally, the bias towards
the reference configuration can be imposed by fixing some of the
degrees of freedom -in practice the position of a fraction of the
particles- to the values they take in the reference configuration
\cite{Kim03,BK12,KB13,BC12,KP13}. This is called the ``pinned
particles'' method.

The three ways of constraining the system have different advantages
and reveal different aspects of metastability. In both the annealed
and quenched potentials metastability is revealed by the shape of the
potential function. It is well known that many aspects of dynamical
Mode Coupling Theory-like transitions \cite{Go08}, including dynamical
heterogeneities and growth of correlations, can be seen studied from
the quenched potential construction
\cite{FP00,DFGP02,FPRR11,FJPUZ13-1,FJPUZ13-2}.  Using the overlap as
an order parameter, as it is done in the annealed and quenched
potential cases, allows to discriminate the thermodynamic view of the
glass transition, where the overlap among configurations is the
relevant order parameter, from a purely kinetic one, where the overlap
does not allow to discriminate different metastable states
\cite{FKZ12,FS13}.

The pinned particles method on the other hand, does not uses directly
the overlap as an order parameter, but is attractive because it 
can add stability to metastable states in a way that
% does not allow this discrimination \footnote{We base this consideration on
%   the observation that in kinetically constrained models such as the
%   Fredickson-Andersen on random graph or similar models that present a
%   dynamical glass transition, the transition is displaced to lower
%   density in the pinned particle method, while it remains at
%   the same density with the potential method \cite{preparation}. 
% %%%SF
% }, however, 
%in the case of {\it symmetric pinning}
%where the temperature of the pinned particles and the one of the free
%ones are the same, has the advantage that 
the equilibrium state of
the system is not perturbed.

A marking feature of glassyness as we know from Mean-Field theory is
the fact that in all three cases, the imposed constraint induces new
phase transitions in the system \cite{FP97,FP98}-\cite{BC12}. The
nature of these transitions differs in the different procedures
\cite{CaBi13}. In the cases of attractive interactions one finds a
first order transition line in the plane of temperature and
interaction strength \cite{FP97,FP98}. In the case of pinned
particles, the constraint induces a line of phase transition in the
plane of temperature and fraction of blocked particles.  Here the nature of
the transitions depends in the detailed procedure of pinning, one can
either find a first order transition line as in the case of the
coupled systems, or instead a line of ideal glass transition with
Kauzmann entropy crisis that crosses over to a line of second order
glass transition \cite{Ca13}.  In all cases, but this last one,
the line of phase transition terminates in a critical point.  The
existence of these critical points is a crucial prediction of the
thermodynamic mean field approach. On the contrary, in purely dynamic
theories of glassyness \cite{GC02} and in exactly solvable kinetically
constrained models \cite{GST11} (such as e.g. the Fredickson-Andersen
on random graph or similar models) the lines of thermodynamic phase
transition and the critical point are not present \cite{JB12}.  Their
existence is therefore one of the few discriminating predictions that
are different between the two approaches.  The thermodynamic scenario
has started to receive confirmations in numerical simulations of
liquid systems. In \cite{CCGGPV10} it was provided evidence for a
coupling induced first order transitions in the quenched potential
setting. In \cite{Berthier} this result was confirmed, both for the
annealed and the quenched and it was convincingly shown that the line
of phase transition terminates in a critical point. Other numerical
results in this sense will be presented soon \footnote{G. Parisi and
  B. Seoane, In preparation}.
% The possibility of
% observing the critical points opens an important window for numerical
% verifications of theoretical predictions of Mean Field theory in
% realistic models.
In this paper we address the problem of the
characterization of the universal properties of these critical
points. Simple arguments can be put forward to understand these
properties.  In the annealed potential case, the only source of
overlap fluctuations is the thermal noise.  The critical point is
described by a quartic field theory and is in the Ising universality
class. In the quenched and pinned cases, however, a second source of
fluctuations can be identified in the choice of the reference
configuration \cite{FPRR11}.  This acts as a random field in the
system and the resulting universality class is the one of the Random
Field Ising Model (RFIM) \cite{BY91,N98}. In order to turn these
qualitative arguments into an accomplished theory a deep analysis of
the soft modes emerging at the critical point and the properties of
the perturbation theory should be performed. Replica Field Theory
(RFT), in terms of which the constrained free-energy can be in
principle computed, provides the natural formal setting to frame the
problem. The three different procedure are found to correspond to
different underlying symmetries and/or analytic continuations in the
number of replicas that one should consider. 

The analysis of perturbation theory of replica field theories
describing glassy criticality has been initiated in \cite{FPRR11}
where it was shown how the description of dynamical heterogeneities in
the beta regime, close to a mode coupling (MCT) dynamical transition
could be mapped in a spinodal point of a RFIM with cubic interaction.
In a subsequent paper \cite{FPR13} it was analyzed the case of a
replica symmetric theory where the leading cubic interaction term
vanishes.  This theory describes higher order glass singularities as
well as the critical point of the {\it symmetric} pinned particle
construction where the the pinned particles are blocked from a
configuration equilibrated at a temperature equal to the one at which
the free particles evolve. In that case the universality class of the
($\phi^4$) RFIM was found within a perturbative one loop
calculation. Here we extend our analysis to the annealed and the
quenched potential and asymmetric pinning where the pinned particles
are blocked from a temperature smaller than the one of the free
particles, that we are able to treat at all orders of perturbation
theory.  In all cases we find that the expectations from the
qualitative argument are met. While the annealed case is attractive
for the simplicity of the result and the possibility to verify it in
numerical simulations, most interesting from the theoretical point of
view, for its implication on the nature of fluctuations and
heterogeinities in glassy systems are the quenched potential case
\cite{FPRR11} and the asymmetric pinning case, where our analysis
shows how the Parisi-Sourlas supersymmetry \cite{PS82} of the RFIM
naturally emerges at criticality.

The plan of the paper is the following: In the next section we shortly
review the theory of glassy systems under constraints. In section 3 we briefly
discuss the annealed case. Then in section 4 we state the problem of
the critical point for the quenched potential and pinned particle
case. We analyze the zero modes of the mass matrix in section 5. In
section 6 we derive the RFIM action by dimensional analysis. In
section 7 we discuss physical correlation functions and their
relations.  We finally summarize and conclude the paper. An appendix
presents some technical details.

\section{Glassy systems under constraints}
In this section we briefly review the use of constraints to unveil
glassyness and metastability. Let us consider a system described by
the Hamiltonian $H(X)$ where $X$ specifies the configuration of all
the particles in the system. We suppose a-priori that there is no
quenched disorder in $H$, even though this could be included. As we
will see, in all cases the computation of the constrained free-energy
can be tackled through the use of the replica method. The order
parameter of the theory is an overlap matrix, and fluctuations will be
described by a Landau expansion of the free-energy around a saddle
point. In the specific the problems differ in the number of replicas,
which is 2 in the annealed potential problem and $n\to 1$ in the
quenched potential and the pinned particle case, and in the symmetry
of the saddle point: $S_{n-1}$ in the quenched potential and
asymmetric pinning, $S_n$ in the symmetric pinning.
\subsection{Annealed Potential Construction}
The simplest setting consists in considering two copies in the system 
interacting through an attraction \cite{FPV92,FP98} 
\begin{eqnarray}
  H_2(X,Y)=H(X)+H(Y)+N \epsilon q(X,Y). 
\end{eqnarray}
where for a system with $N$ particles, 
the overlap $q(X,Y)$ among two configurations $X=\{x_1,...,x_N\}$ and 
$Y=\{y_1,...,y_N\}$ can be defined in terms of a short range 
attractive interaction potential $w(x)$ as
\begin{eqnarray}
\label{eq:ove}
q(X,Y)=\frac 1 N \sum_{i,j}w(x_i-y_j). 
\end{eqnarray}
Space dependent overlap fields $q(x;X,Y)$ can be defined restricting
the sum in (\ref{eq:ove}) to the particles in some neighborhood of $x$. 

The free-energy of the system $F(\epsilon,T)$ involves a sum over the
configurations of the two copies -or replicas- of the system.  One can
see this sum as a particular case of a replicated system where the
number of replicas $n$ here is just equal to 2, and, just as in the
case of uncoupled systems, the study of liquid phases can be addressed
without need for analytic continuations. Conversely, the study of
glassy phases requires analytic continuations in the number of
replicas, but since, as we will see, the critical point we are
interested lies in a liquid region we will not need to consider these
continuations.

The Legendre transform of $F(\epsilon,T)$,
$W(q,T)=F(\epsilon,T)+\epsilon q$ is called annealed potential
function. We refer broadly to this procedure of symmetric coupling as
annealed potential construction.

\begin{figure}
\label{figura}
\epsfxsize=220pt 
\epsffile{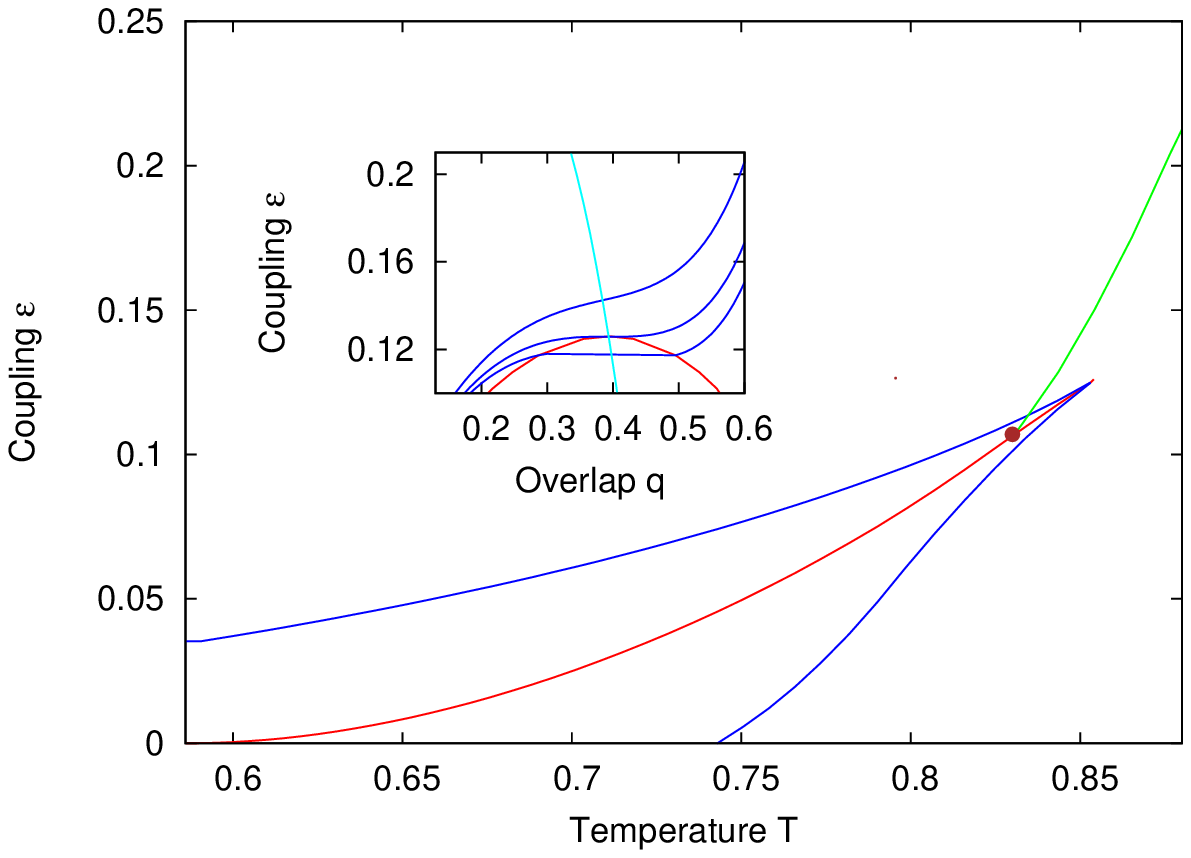}
\epsfxsize=220pt 
`\epsffile{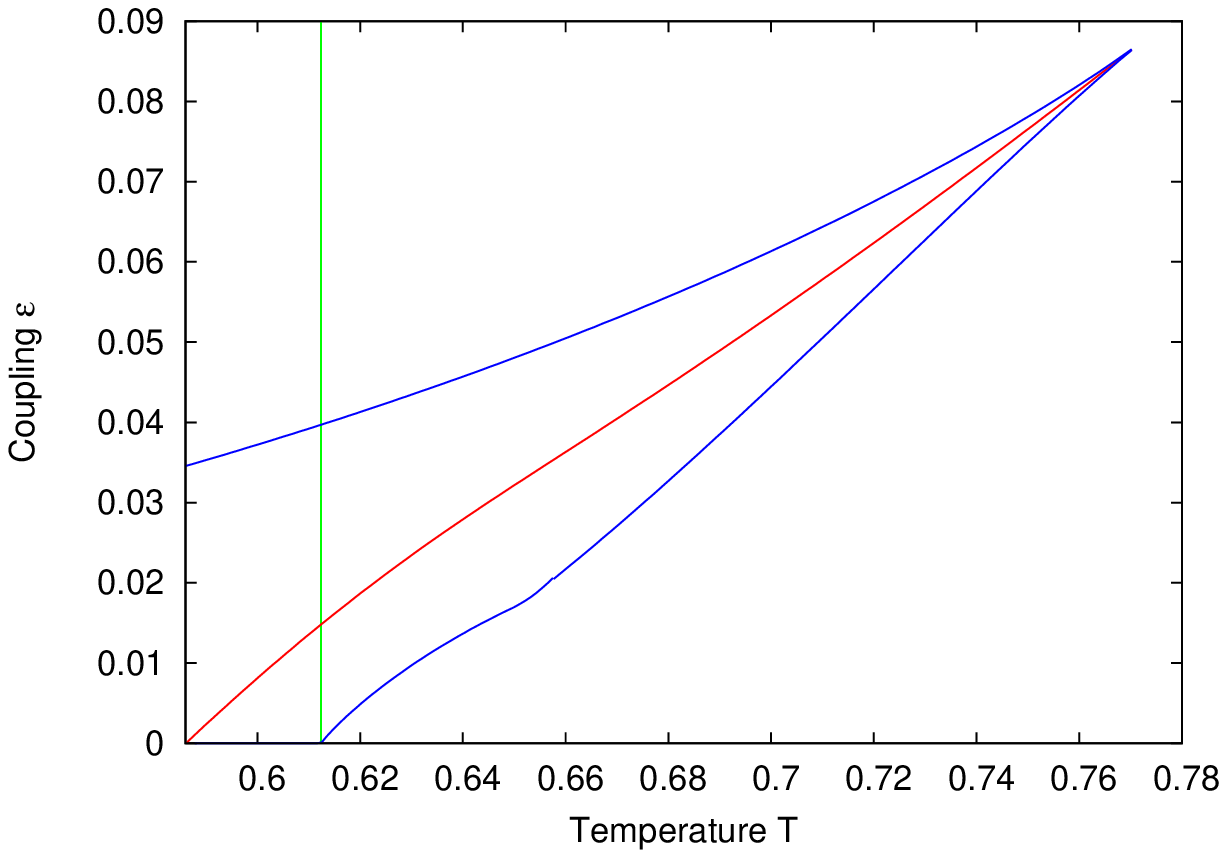}
\caption{Typical Mean Field phase diagram in the $T - \epsilon$ plane
  for the annealed and the quenched constructions.  {\bf Left panel}: The
  annealed construction phase diagram.  The central (red) line marks
  the first order phase transition between a confined high overlap
  phase above and an unconfined phase below.  The green line is the
  line of dynamical (MCT-like) glass transition. The confined phase is
  a glass to the left of the brown point where the red line and the
  green line meet and it is a liquid to its right. The blue lines are
  the spinodal lines of the confined (lower curve) and unconfined
  (upper curve) phases.  In the inset we show the typical isothermal
  and coexistence lines in the $q-\epsilon$ plane.  From top to bottom
  we have an isothermal in the single phase region $T>T_{Cr}$, the
  critical isothermal $T=T_{Cr}$ and a isothermal in the two phase
  region $T<T_{Cr}$, the horizontal line corresponds to Maxwell
  construction. The cyan line is the Widom line: the locus of points
  where the potential has an inflection and $g_3 =0$ and the
  susceptibility $\chi_4=d\langle p\rangle/d\epsilon$ has a maximum.
  {\bf Right panel}: Phase diagram of the quenched construction The
  transition line and its corresponding spinodal are similar to the
  annealed case, however here the coupling does not induces new glass
  transitions.  Glassyness appears at the dynamical glass transition
  temperature $T_d$ of the unconstrained model (vertical green
  line). Notice the different scales in the two panels. The quenched
  critical point lies at lower temperature and coupling than the annealed one. 
We have used here the spherical $p$-spin
  model \cite{CS92} for $p=3$ for which the dynamical (MCT) transition
  temperature is $T_d=0.612$ and the Kauzmann transition temperature
  is $T_k=0.586$.  }
\label{fig1}
\end{figure}

In Mean Field models with a glass transition, like e.g. p-spin or
Potts spin glasses\cite{FP95,Me99} or liquids in the HNC approximation
\cite{CFP99} the coupling induces temperature dependent phase
transitions in the system. A typical phase diagram is presented in
figure \ref{figura} In the temperature-coupling plane, one finds both
a line of ordinary liquid-glass transition where the overlap between
the two copies is non singular \cite{Me99}, and a line of first order
phase transition that separates a low overlap or deconfined, phase,
where the two copies are weakly correlated, from a high overlap or
confined phase where the two replicas stay close to each other. The
first order transition line, which departs from the ideal Kauzmann
transition temperature $T_k$ for $\epsilon=0$, terminates in a
critical point $(T_{Cr},\epsilon_{Cr})$. Interestingly, this line and
the glass transition line meet in a point, and while the deconfined
phase is always a liquid, depending on the temperature, the confined
phase can be either a liquid or a glass.  What is important for us is
that a whole part of the line, which includes the critical point,
marks the border of a liquid-liquid transition. The critical point
lies at a finite distance from the line of glass transition and a
description with just two replica is appropriate.

The inset of figure \ref{figura} shows the typical isothermal lines
and coexistence curve in the overlap-coupling plane for temperatures
close to the critical point. Notice the similarity to the isothermal
of the gas-liquid phase transition in the $V-p$ plane.  As in this
case, the critical fluctuations  can be described
expanding the free-energy with respect to the the local fluctuation of
the order parameter around its (space homogeneous) average value
$q^*$. We can notice the similarity of figure \ref{figura} with the
coexistence diagram recently obtained in numerical simulations of a
realistic liquid model by Berthier \cite{Berthier}.

\subsection{Quenched Potential Construction}
The symmetric coupling between replicas in the annealed procedure
introduces strong biases to the equilibrium. In order to faithfully
explore the vicinity of typical equilibrium states at temperature, one
considers instead a quenched procedure where one fixes a reference
configuration $X_0$ extracted with Boltzmann probability at a
temperature $T_{ref}$, and uses the particle positions in the reference
configuration to define an external potential in which the particles
of the constrained system $X$ evolve \cite{FP95,FP97,FP98}.  The
Hamiltonian of the system for fixed $X_0$ is
\begin{eqnarray}
  H_\epsilon (X)=H(X)+N \epsilon q(X,X_0). 
\end{eqnarray}
It should be noted the fundamental asymmetry between $X_0$, which is
just a random equilibrium configuration and the system $X$ which feels
an attraction towards $X_0$. The reference configuration can be
considered as a sort of quenched disorder in which the system
evolves. The constrained free-energy $F_Q(\epsilon,T,T_{ref})$ is defined as
\begin{eqnarray}
  F_Q(\epsilon,T,T_{ref})=-\frac T N \frac {1} {Z(T_{ref})} \sum_{X_0} e^{-\beta_{ref} H(X_0)} \log \left(\sum_X  e^{-\beta (H(X)+N \epsilon q(X,X_0))}\right). 
\end{eqnarray}
One usually chooses $T_{ref}=T$, but the case in which the temperature
of the system is different from the temperature has also been
considered \cite{FP98,KrZd10} to study the evolution of metastable
states with temperature.  The quenched average over the distribution
of the reference configuration is usually dealt with the replica
method. One needs to replicate the system $X$ a number $n'$ of times
and perform a continuation $n'\to 0$ at the end of the
computations. Noticing that the reference configuration can be seen as
an additional replica, the total number of replicas is $n=1+n'$ which
should be sent to 1. Due to the asymmetry in the interaction between
reference configuration and the system, replica number ``0'' turns out
to be privileged. In fact the effective Hamiltonian reads:
\begin{eqnarray}
H_{eff}(X_0,X_1,...,X_{n-1})=\sum_{a=0}^{n-1} H(X_a)+\epsilon 
\sum_{a=1}^{n-1}q(X_0,X_a)  
\end{eqnarray}
(notice that the index $a$ runs over different ranges in the two
sums).  Instead of possessing the familiar symmetry $S_n$ under
permutations of all the $n$ replicas, the problem is symmetric only
under the permutations $S_{n-1}$ of replicas with index $a>0$.
Analogously to the annealed case, mean field theory predicts the
existence of a line of phase transition in the $\epsilon - T$ plane
that terminates in a critical point \cite{CFP97,CFP99,FP98} and
separate a confined phase with high overlap with the reference
configuration from a deconfined phase with low overlap.

\subsection{Particle pinning}
Just as in the quenched potential case, in the case of particle
pinning one fixes a reference configuration in $X_0$ from the
equilibrium distribution at a temperature $T_{ref}$, but then one
considers configurations $X$ in which a fraction $\theta$ of the
variables are fixed to the values they take in $X_0$
\cite{Kim03,BK12,KB13,BC12,KP13}. Also in this case Mean Field Theory
predicts that the reduction of degrees of freedom induces new phase
transitions in the system. Interestingly, the nature of the phase
transitions depends on the details of the pinning procedure. As
discussed in much detail in \cite{Ca13}, if $T_{ref}=T/\alpha$ with
$\alpha > 1$ ($T_{ref}<T$) one finds a pattern of phase transition
similar to the one of the annealed and quenched potential, there is
line of confinement first order phase transition in the $\theta - T$
plane that terminates in a critical point. Conversely, if $\alpha<1$
($T_{ref}>T$) one finds a line of ideal RFOT Kauzmann-like transition
of the discontinuous 1RSB kind that crosses-over into a line of second
order glass transition of the continuous 1RSB kind
\cite{BC12,Ca13}. The nature of the terminating point of the first
order transition in the first case and the RFOT transition in the
second case is rather different as we will discuss in the next
section.

Within the replica method this
procedure still requires an analytic continuation in the number of
replicas $n$ which tends to 1. If the temperature of the reference
configuration $T_{ref}$ is different from the temperature of the
non-pinned particles, still the reference configuration is singled out
and the symmetry is $S_{n-1}$. In the important case $T_{ref}=T$
however, one can show that the unpinned particles remain at
equilibrium \cite{FS11,Kr10}\footnote{The line $T_{ref}=T$ can be seen
  as a symmetric line analogous to the Nishimori line familiar in spin
  glass theory \cite{KZ11}.}. As a consequence, within the replica
formalism there is full $S_n$ replica symmetry. The problem of
critical point in $n\to 1$, $S_n$ symmetric replica field theories has
been addressed in \cite{FPR13}. In that case through the analysis of
the soft modes of the replica field theory the critical point was
shown to belong to the RFIM universality class. In this paper we
extend our analysis to the case of $n\to 1$, $S_{n-1}$ symmetric
theories.

\section{The Annealed critical point}
As we stated in the previous section, in the annealed potential case
if we describe liquid phases, the complexity of the replica method is
reduced to minimal terms. There are just two replicas and the $n\times
n$ overlap order parameter matrix $q_{a,b}$ which appears in the
replica method has here a single independent entry $q$ with $a\ne b$.
This represents the overlap between the two copies, it is the only
order parameter of the problem. \footnote{ Mean-field theory is based
  on models with quenched disorder of the family of the spherical
  p-spin model. There the so called ``annealed approximation'' where
  the average partition function rather that the average free-energy
  is evaluated turn out to be exact and allows to compute exactly the
  potential.}  
The resulting Landau free-energy is a functional of a
single field $\phi(x)=q(x)-q^*$ representing the fluctuation of the
overlap around its average,
\begin{eqnarray}
  \label{eq:Landau}
&&  F[\phi]=\int dx\; \frac 1 2 k \; (\nabla \phi(x))^2+V(\phi(x))\nonumber
\\
&& V(\phi)= \frac 1 2  m_0\; \phi^2 +g_3 \; \phi^3 + g_4 \; \phi^4. 
\end{eqnarray}
The coefficients $m_0$, $g_3$, $g_4$ as well as $q^*$ smoothly depend
on the control parameters $T$ and $\epsilon$. Away from the critical
point, where $g_3\ne 0$ the quartic term is irrelevant in perturbation
theory, however, as in the gas-liquid transition case, at the critical
point, both $m_0$ {\it and} $g_3$ vanish. We find therefore that the
critical point is described by an ordinary scalar field theory with
$\phi^4$ interaction and is in the universality class of the ordinary
Ising model. This is coherent with the recent analysis of \cite{Sho12,DJS13}. 

\section{The Quenched and Pinned Critical Points} 
In order to be defined, we consider the context of quenched potential
however, the main ingredients within Replica Field Theory being the
number replicas and the relative symmetry, with little modifications, that
we will specify on the way,  
one can treat the pinned particle construction with $T_{ref}<T$.  %This is the
%case for simple liquid systems in the HNC approximation
%\cite{CFP97,CFP99} and it is the scenario found quite generically in
%mean-field spin glasses \cite{FP98}. 
Replicas can be used to
average over the choice of the reference configuration and we would like
to describe the class of universality of the critical point within
replica field theory. The starting point will be a replica field
theory with $n\to 1$ replicas over a space dependent $n\times n$ space
dependent matrix $Q_{ab}(x)$ of the kind:
\begin{eqnarray}
  F[Q_{ab}(x),\epsilon]=F_0[Q_{ab}(x)]-\epsilon \sum_{a=1}^{n-1} 
\int dx \; Q_{0a}(x). 
\end{eqnarray}
The term $F_0[Q_{ab}(x)]$ is symmetric under permutation of all
replicas. The last term in the action breaks this symmetry, in fact
replica number 0, which corresponds to the reference state is
privileged with respect to the others. The symmetry $S_{n-1}$ under
permutations of replicas $1,...,n-1$ remains unbroken. In the pinning
particle construction the $\epsilon$-coupling term is absent, however,
if the temperature of the reference configuration $T_{ref}$ is
different from the temperature $T$ at which the free particles evolve,
the reference configuration is singled out and again $F[Q_{ab}(x)]$ 
contains terms that break $S_n$ into $S_{n-1}$. 

As usual we will start from a Landau
expansion of the free-energy close to the critical point, supposing
that the non-diagonal elements of the matrix $Q_{ab}(x)$ can be
written as
\begin{eqnarray}
Q_{ab}(x)=  Q_{ab}^*+\phi_{ab}(x)
\end{eqnarray}
where $Q_{ab}^*$ is the saddle point value of the matrix order
parameter. The diagonal elements, which are related to the structure 
factor, in general  are non critical, regular across the transition point and 
will not be discussed here. 

The saddle point matrix $Q_{ab}^*$ is homogeneous in space and for the
critical points we consider here has the replica symmetric form
$Q_{a,b}^*=q+(\delta_{a0}+\delta_{b0})(p-q)$. The parameters $p\ne q$
represent respectively the overlap between the system and the
reference configuration and the self-overlap of the system with
itself. Instead of trying to write the most general
$S_{n-1}$-invariant polynomial expansion of the free-energy in terms
of $\phi_{ab}(x)$, our strategy will consist first to analyze the
properties of a $S_{n-1}$ invariant mass matrix close to criticality
and then after identified the soft modes, in writing directly the
generic field theory describing their interaction disregarding
completely the massive modes. As a preliminary let us study
longitudinal fluctuations, i.e. just fluctuations of $p$ and $q$.  At
the saddle point level, the free-energy as a function of $\epsilon$
reads
\begin{eqnarray}
\label{eq:pot}
  && \Gamma [\epsilon]=\frac{\partial}{\partial n}|_{n=1}  F[Q_{ab}^*,\epsilon]\\
  &&  \Gamma [\epsilon]=N\left( W[p,q]-\epsilon p \right) \nonumber\\
  && \frac{\partial W}{\partial q}=0\;\;\;\;
  \frac{\partial W}{\partial p}=\epsilon. \nonumber
\end{eqnarray}
 As usual one can interpret the
effective potential $V(p,\epsilon)=W[p,q(p)]-\epsilon p$ at the point
$q(p)$ defined by $\frac{\partial W}{\partial q}=0$ as the value of
the free-energy when the system to reference configuration 
overlap takes the value $p$. The physical
value of $p$ is fixed by the stationary condition $V'[p]=0$.  At a
critical point terminating a first order line one should have in
addition that the second and the third derivatives of $V$ vanish,
$V''[p]=V'''[p]=0$, conditions that generically fix the values of $T$
and $\epsilon$.  Let us remark that close to a generic saddle point
values, away from the critical point, 
the function $W[p,q]$ must admit an expansion of the kind:
\begin{eqnarray}
\label{eq:expansion}
  W[p+\delta p,q+\delta q] -W[p,q]= &&\frac 1 2 \left[ \hat{M}_{pp} \delta p^2 + 2  \hat{M}_{pq} \delta p \delta q +  \hat{M}_{qq} \delta q^2 \right]
  \nonumber\\
  && +\sum_{r=0}^3 C_r \delta p^r \delta q^{3-r} +O(p^4)
\end{eqnarray}
This function describes longitudinal fluctuations which are constant
in space and the form of $\phi$ is the same as the one of the saddle
point. By definition, at the critical point, longitudinal fluctuations
for which $\delta q=\frac{dq(p)}{dp}\; \delta p=- \hat{M}_{pq}/
\hat{M}_{qq}\; \delta p +O(\delta p^2)$ are long ranged, the quadratic
and the cubic forms vanish and the quartic terms become important. Of
course, the points where the quadratic form vanishes but the cubic
form remains finite are also critical. These correspond rather to
spinodal points than to thermodynamic critical points.  A well known
example is the one of dynamical glass transitions points that
correspond to $S_n$ symmetric cubic theories
\cite{FPRR11,FJPUZ13-1,FJPUZ13-2}. In the present case, as it can be seen in
fig. \ref{figura} there are just two spinodal lines, for the confined
and the deconfined phases that converge into the critical point for
$(T,\epsilon)\to (T_{Cr},\epsilon_{Cr})$. Our analysis shows that
generically, despite their different nature \cite{CaBi13}, both these
spinodals belong to the $\phi^3$-RFIM universality class.

If one considers longitudinal fluctuations that are not constant in space, an 
additional ``kinetic term'' of the kind
\begin{eqnarray}
  K[\delta p(x),\delta q(x)]=\frac 1 2 \int dx\; [k_p (\nabla \delta p)^2+k_q(\nabla \delta q)^2]
\end{eqnarray}
is present in the longitudinal Landau expansion. 

Close to the critical point, the mass of the soft
mode can be simply related to the coefficient to the quadratic form, to the 
lowest order, 
\begin{eqnarray}
  \hat{m}_0=\frac{ \hat{M}_{pp} \hat{M}_{qq}- \hat{M}_{pq}^2}
{ \hat{M}_{pp}+ \hat{M}_{qq}} + O(\hat{m}_0^2)
\end{eqnarray}

\section{The mass matrix and its eigenspaces}
In this section we would like to go beyond longitudinal fluctuations,
and identify all the zero modes of the problem in order to build up
the suitable critical theory that describes their interaction.
The physical meaning of the relevant modes will result from their contribution 
to the various kinds of correlation functions that we discuss in section 
\ref{sec:corr}. 

 Let us study the most general mass matrices of small
fluctuations, actually a 4-index ``matroid'', $M[a,b;c,d]$ that is
symmetric under the operations $(ab)\to (ba)$, $(ab;cd)\to (cd;ab)$,
vanishes if $a=b$ or $c=d$ and respects the $S_{n-1}$ replica
symmetry.  Such a matrix has at most 7 distinct elements that can be
parametrized in the following way: (all indexes are assumed to be
different among themselves and different from $0$ in the next
formulae)
\begin{eqnarray}
&  M[0,a;0,a]=\frac{m_1}{2}+\frac{\mu_2}{2}+\mu_3;
\;\;\; &M[0,a;0,b]=\frac{\mu_2}{4}+\mu_3
\nonumber \\
&
M[0,a;a,b]=\frac{\nu_2}{4}+\nu_3;
\;\;\;
&M[0,a;b,c]=\nu_3
\nonumber \\
&
M[a,b;a,b]=\frac{m_1}{2}+\frac{m_2}{2}+m_3;
\;\;\;
&M[a,b;a,c]=\frac{m_2}{4}+m_3
\nonumber \\
&
M[a,b;c,d]=m_3&
\nonumber
\end{eqnarray}
The parameters $m_1,\; m_2,\; m_3,\; \mu_2,\; \mu_3,\; \nu_2,\; \nu_3$
can be supposed to be distinct.
Notice that the usual $S_n$ symmetric matrix is recovered if one poses
$\mu_2=\nu_2=m_2$ and $\mu_3=\nu_3=m_3$. 

We now look at the eigenspaces of $M$ proceeding analogously to the
classical De Almeida-Thouless analysis of fully $S_n$ invariant
matrices \cite{AT78}. In full generality the eigenspaces of $M$ can be
related to the representation of $S_{n-1}$ over symmetric (two index)
matrices with vanishing diagonal elements. These representations are
well known and consist in replica symmetric matrices, matrices that
break the symmetry privileging one replica and matrices that privilege
two replicas. In the following we use the terminology usually employed
in spin glass theory, calling respectively Longitudinal the Anomalous
and Replicon these eigenspaces.
%sectors that are
%respectively $S_{n-1}$ symmetric or they privilege one or two
%replicas.

\subsection{The longitudinal space}
The simplest eigenvector are the longitudinal ones
that have the same structure of the saddle point $Q_{ab}$, for $a\ne b$:
\begin{eqnarray}
\label{eq:long}
L_{ab}=(u-v)(\delta_{a0}+\delta_{b0})+v  
\end{eqnarray}
to which there correspond the two eigenvalues $\lambda_{LO}^\pm$ that
are given in the appendix.  One of these, that we call $\lambda_{LO}$
vanishes at the critical point while the other remains finite. Notice
that $u$ and $v$ can be identified respectively with the variations
$\delta p$ and $\delta q$ of the previous section.  Comparing the
quadratic form $\langle
L|M|L\rangle=\sum_{ab,cd}L_{ab}M[a,b;c,d]L_{cd}$ in the limit $n\to 1$
with the one appearing in (\ref{eq:expansion}), one can identify $
\hat{M}_{pp},\; \hat{M}_{qq}$ and $ \hat{M}_{pq}$ as
 \begin{eqnarray}
   \label{eq:iden}
   \hat{M}_{pp}= 2 m_1+\mu_2,\;\;\;  \hat{M}_{qq}= m_2-m_1,\;\;\;  \hat{M}_{pq}= -\nu_2
 \end{eqnarray}
Strictly speaking the longitudinal eigenvalues of the matrix $M$ do
not coincide with the eigenvalues of the quadratic form in
(\ref{eq:expansion}). This due to the fact that 
if we write $L_{ab}=u w^1_{ab} + v w^2_{ab}$ 
the vectors $w^1$ and $w^2$ are not normalized to
1. However, it is easy to see that if one of the eigenvalue of the
quadratic form is zero, so it is the corresponding eigenvalue of $M$. 
Simple linear algebra shows that for $n\to 1$ the small longitudinal eigenvalue 
$\lambda_{LO}$ 
reads
\begin{eqnarray}
  \label{eq:lin}
  \lambda_{LO}|_{n=1 }\equiv m_0 = \frac{\hat{M}_{pp}+\hat{M}_{qq}}{2\hat{M}_{qq}-\hat{M}_{pp}}\hat{m}_0 +O(m_0^2). 
\end{eqnarray}
The corresponding eigenvector should be such that for $n\to 1$, $v=
\frac{dq(p)}{dp} u=- \hat{M}_{pq}/ \hat{M}_{qq} u$. For future
reference we introduce the notation $\gamma= \frac{dq(p)}{dp}$. In
general it can be expected $\gamma>0$, implying strong correlations 
between the fluctuations of $p$ and these of $q$. 

\subsection{The anomalous space}

The second family of eigenvectors are the so-called anomalous
ones $a_{ab}^\mu$, where in addition to replica 0 a replica $\mu>0$ is
privileged: there are $4$ distinct elements (all indexes are different among themselves and from 0):
\begin{eqnarray}
&&  a^\mu_{0\mu}=u_0+u_1;
\;\;\;
a^\mu_{0a}=u_1
\nonumber
\\
&&
a^\mu_{\mu a}=v_0+v_1;
\;\;\;
a^\mu_{ab}=v_1.
\end{eqnarray}
% or, using $\delta$ functions: 
% \begin{eqnarray}
%   A^\mu_{ab}=[\delta_{a0}(\delta_{b\mu}u_0+u_1)+\delta_{b0}(\delta_{a\mu}u_0+u_1)]+
% (1-\delta_{a0}-\delta_{b0})[(\delta_{a\mu}+\delta_{b\mu})v_0+v_1]
% \end{eqnarray}
If we impose orthogonality between the anomalous and longitudinal
spaces we find:
\begin{eqnarray}
  && u_0=-(n-1)u_1\\
  && v_0=-\frac 1 2 (n-1) v_1
\end{eqnarray}
This fixes two parameters out of four and also in this case there are
two independent eigenvalues. Notice that $u_0$ and $v_0$ that are
responsible for the difference between $a^\mu$ and $L$, are of order
$n-1$ relative to $u_1$ and $v_1$, this implies, on a very general
basis that the anomalous and longitudinal eigenvalues form degenerate
doublets for $n\to 1$. Their difference which should be linear in
$u_0$ and $v_0$ is of the order $n-1$.
The values of $v_1$ and $u_1$ become degenerate with the values of the
parameters $v$ and $u$ in the corresponding longitudinal eigevectors.
This is confirmed by the explicit computation of eigenvalues and
eigenvectors as a function of the mass matrix parameters with
Mathematica.  As we will see this eigenvalue degeneracy is at
the origin of typical random field terms in the action. 

The total
dimension of the anomalous space is $2(n-2)$ as it can be realized
taking into account the orthogonality with the longitudinal space.

The meaning of the anomalous vectors can be understood within the 
replica formalism looking at the 
projection of the fluctuating field: 
\begin{eqnarray}
\langle \phi|a^\mu \rangle =&& u_1[-(n-1)\phi_{0\mu}+\sum_{b=1}^{n-1} \phi_{0b}]
\nonumber\\&&+
v_1 [-(n-1) \sum_{a=1}^{n-1}\phi_{\mu a}+\sum_{a,b=1}^{n-1}\phi_{ab}]. 
\end{eqnarray}
These are replica symmetry breaking fluctuations where of
$\phi_{0\mu}$ and $\sum_{a=1}^{n-1}\phi_{\mu a}$ differ from their
averages over the index $\mu$. Soft modes in these directions have as
physical consequence deep relations among different correlation
functions that can be defined, as we discuss in section
\ref{sec:corr}.

\subsection{The replicon space}
The last family of eigenvectors is the one of so-called
replicons. These can be characterized in two equivalent ways: either
as matrices that besides replica number ``0'' privilege two replicas
$\mu,\nu>0$ and are orthogonal to the longitudinal and anomalous
spaces, or, more simply, as matrices $R_{ab}$ such that
\begin{eqnarray}
R_{a0}=0;\;\;\;\; 
\sum_{b} R_{ab}=0\;\;\; \forall \;\; a.
\end{eqnarray}
% One can see that they read:
% \begin{eqnarray}
%   R^{\mu\nu}_{ab}&=&[\delta_{a0}((\delta_{b\mu}+\delta_{b\nu})
% r_0+r_1)+
% [\delta_{b0}((\delta_{a\mu}+\delta_{a\nu})
% r_0+r_1)
% ]+
% \nonumber\\ &&
% (1-\delta_{a0}-\delta_{b0})
% [
% (\delta_{a\mu}+\delta_{a\nu}+\delta_{b\mu}+\delta_{b\nu})s_0+
% (\delta_{a\mu}\delta_{b\nu}+\delta_{b\mu}\delta_{a\nu})s_1+
% s_2
% ].
% \end{eqnarray}
The replicon space concerns fluctuations of the $\phi_{ab}$ which are
independent from the ones of $\phi_{0b}$ and induce replica symmetry
breaking of the type familiar from spin glass theory \cite{MPV}. The
dimensionality of the replicon sub-space is $(n-1)(n-4)/2$. Together
with the longitudinal and anomalous spaces it exhausts the $n(n-1)/2$
dimensional linear space of symmetric matrices null on the diagonal.
It is easy to see, using the form of the mass matrix that there is a
single replicon eigenvalue and it is just given by $m_1$.  The
behavior of the replicon eigenvalue marks the difference between
critical points terminating the first order transition lines of the
quenched construction and the pinned particles one for $T_{ref}<T$ and
the critical points marking the passage from a RFOT Kauzmann (or
discontinuous 1RSB) transition to a continuous 1RSB glass transition
for the pinned particles construction with $T_{ref}>T$.  Generically,
in the former case, $m_1$ does not have reasons to vanish at the
critical point. For example it can be checked that $m_1$ indeed
remains finite at the critical point of the spherical p-spin model in
the quenched potential setting.  Conversely, in the latter case $m_1$
vanishes at the transition and is zero on the whole second order glass
transition line. In this paper we concentrate on the case that $m_1$
remains positive at criticality, and treat therefore the terminating
critical points of the first order lines of the quenched construction
and the pinned particle problem with $T_{ref}<T$. The
case $T_{ref}= T$ where full $S_n$ replica symmetry is recovered marks
the boundary between the two behaviors. This case has been analyzed
within a one loop approximation in \cite{FPR13}. The analysis
performed there shows that the additional degeneracy of the small
eigenvalue does not change the universality class of the problem. In
the replica formalism, we think this could be related the
peculiarities of the $n\to 1$ limit and one can conjecture that the
RFIM universality class also holds for $T_{ref}>T$. Further work will
be needed to extend the analysis of \cite{FPR13} to all orders in
perturbation theory.  In conclusion, generically, at critical points
terminating first order lines, the singularities come from the fact
the longitudinal and anomalous fluctuations go soft at the transition,
while the replicon ones remain massive.
%Notice that this is at variance with the usual spin glass
%transition where the replicon modes are the critical ones \cite{AT78}.

We notice that an alternative approach to analyze the eigenspaces of
$M$ (that leads to same results) consists in treating separately the
first line and column of the matrix $\phi_{a,b}$ $a\ne b =0,...,n-1$
as a $n-1$ dimensional vector $z_a=\phi_{0a}$ in replica space with
$a=1,...,n-1$ and the remaining part of the $\phi_{ab}$ matrix with
$a\ne b =1,...,n-1$ as a $n-1 \times n-1$ matrix. In this way the
problem becomes formally closer to the one studied in the paper of De
Almeida and Thouless \cite{AT78}.

\section{A vectorial representation}
To build up the relevant critical theory we can disregard massive
directions and consider an interacting field theory for fluctuating
fields $\phi_{ab}(x)$ which are linear combinations of the critical
modes. We therefore concentrate on the zero mode subsector of the
longitudinal and anomalous spaces and ignore the sectors corresponding
to hard modes. In order to have theory that keeps explicitly the
$S_{n-1}$ symmetry it is convenient to combine longitudinal and
anomalous vectors $L$ and $a^\mu$, that we suppose to be defined up to
a normalization to be fixed a-posteriori, into vectors $A^\mu$
\begin{eqnarray}
A^\mu=a^\mu+L 
\end{eqnarray}
that form an orthonormal basis ${\langle A^\mu|A^\nu\rangle}=\sum_{ab}
A^\mu_{ab} A^\nu_{ab}=\delta_{\mu\nu}$.  Let us state a few
properties of these vectors and fix the normalizations of $a^\mu$ and
$L$.  Notice that while replica symmetry implies that $\sum_\mu a^\mu$
should be proportional to $L$, the orthogonality condition $\langle
a^\mu |L \rangle=0 $ says to us that $\sum_\mu a^\mu=0$. Using again
the replica symmetry we can write
\begin{eqnarray}
 {\langle a^\mu|a^\nu\rangle}
=\delta_{\mu\nu}(\alpha_{11}-\alpha_{12})+\alpha_{12}. 
\end{eqnarray}
Summing over $\mu$ we obtain that $\alpha_{12}=-\frac{\alpha_{11}}{n-2}$. 
Imposing orthonormality we have
\begin{eqnarray}
  &&  {\langle A^\mu|A^\nu\rangle}=\alpha_{12}+ {\langle L|L\rangle}=0\;\;\; \mu\ne\nu\nonumber\\
  &&  {\langle A^\mu|A^\mu\rangle}=\alpha_{11}+ {\langle L|L\rangle}=1
\end{eqnarray}
which implies $\alpha_{11}=1 - {\langle L|L\rangle}=(n-2) {\langle
  L|L\rangle}$ and ${\langle L|L\rangle}=\frac{1}{n-1}$. 

We can now evaluate the matrix element ${\langle
  A^\mu|M|A^\nu\rangle}$.  If we write the longitudinal eigenvalue as
$\lambda_{LO}=m_0+(n-1)\eta_{LO}$ and the anomalous as
$\lambda_{AN}=m_0+(n-1)\eta_{AN}$ and define
$\eta=\eta_{LO}-\eta_{AN}$, we can easily see that
\begin{eqnarray}
 {\langle A^\mu|M|A^\nu\rangle}=m_0\delta_{\mu\nu}+\eta+O(n-1).
\end{eqnarray}
Let us now expand the critical field on the basis of the $A^\mu$
\begin{eqnarray}
\phi_{ab}(x)=\sum_{\mu=2}^n \psi_\mu(x) \; A_{ab}^\mu.  
\end{eqnarray}
We can now formulate the critical theory in terms of the single index
fields $\psi_\mu$. This theory should of course be invariant under all
permutation of the indexes $\mu=1,...,n-1$. Generically the action of
the theory could be written as a sum of a local term which is a polynomial in the fields, and a kinetic term sensitive to space fluctuations of the $\psi_\mu$. 
We are led then to the study of the
replica symmetric low order local polynomial invariants of the $n-1$
component vector of the $\psi_\mu$. These can be build up explicitly,
starting from the monomials of lower orders:
\begin{itemize}
\item 
 The only linear invariant is
\begin{eqnarray}
  I_1=\sum_\mu \psi_\mu.
\end{eqnarray}
\item 
The quadratic invariants are: 
\begin{eqnarray}
 I_{2,1} =I_1^2,\;\;\;\;  I_{2,2} \equiv J_2=\sum_\mu \psi_\mu^2\nonumber
\nonumber
\end{eqnarray}
\item The cubic ones are:
\begin{eqnarray}
I_{3,1} =I_1^3,\;\;\;  I_{3,2}=I_1 I_{2,2},\;\;\;
I_{3,3} \equiv J_3=\sum_\mu \psi_\mu^3.\nonumber 
\nonumber
\end{eqnarray}
\end{itemize}
The higher order invariants can be obviously generated in a recursive
way.  In general, the only invariant of order $k$ which can not be
expressed as a product of lower order ones is: $J_k=\sum_\mu
\psi_\mu^k$. In addition to purely local invariants, we should
consider the lowest order invariant in $\nabla \psi_\mu$, namely, the
kinetic term
\begin{eqnarray}
  \label{eq:kin}
  K[\psi_\mu]=\sum_\mu (\nabla \psi_\mu)^2.
\end{eqnarray}

\section{Dimensional analysis}

The quadratic form ${\langle \phi|M|\phi \rangle}$ 
in the bases of the $\psi^\mu$ is readily computed: 
\begin{eqnarray}
\label{eq:qf}
{\langle \phi|M|\phi \rangle}=m_0\sum_\mu \psi_\mu^2 +\eta
\left(\sum_\mu \psi_\mu\right)^2=m_0 I_{2,2} +\eta I_1^2.  
\end{eqnarray}
We can remark at this point that (\ref{eq:qf}) has the typical form
that appears in the replica treatment of the RFIM with random field
$\delta$-correlated in space. The coefficient $\eta$, that here
originates corresponds in that case to the variance of the random
field. Its appearance here stems from the degeneracy of the
longitudinal and replicon eigenvalues for $n\to 1$. It can be checked in 
specific problems that while $m_0\to 0$ the value of $\eta$ remains positive. 

It is well known from the theory of the RFIM that the
inversion of the form (\ref{eq:qf}) has single pole and double pole
propagators. The field $\psi^\mu$ cannot have a well defined scaling
dimension. The same conclusion can be reached observing that this is
incompatible with the fact that the two terms in eq. (\ref{eq:qf}) are
of the same order of magnitude for $m_0\to 0$ and $\eta$ finite.

In order to use fields with well defined scaling 
dimension we can make a further change of basis
as originally suggested by Cardy \cite{Ca83} and write:
\begin{eqnarray}
\label{base}
&&  \psi_\mu=\psi+\delta_{\mu 1}\hat{\psi}+\chi_\mu\nonumber\\
&& \chi_1=0\nonumber\\
&& \sum_\mu\chi_\mu=0. 
\end{eqnarray}
This is a legitimate change of basis since for all $x$ it contains $n-1$
independent parameters. 
We notice that in this basis the field $\phi$ can be written as 
\begin{eqnarray}
  \phi&&=(n-1)\psi\; L+\hat{\psi} \;A^1+\sum_\mu \chi_\mu\; A^\mu\nonumber\\
&&=[(n-1)\psi+\hat{\psi}]\;L +\hat{\psi}\; a^1+\sum_\mu \chi_\mu\; a^\mu 
\end{eqnarray}
Purely longitudinal fluctuations correspond to
$\hat{\psi}=\chi_\mu=0$. Notice that in this case, the invariants
$J_k$ are of order $n-1$, (in fact $J_k=(n-1)\psi^k$) while all
composite invariants are of higher order. Since the effective
potential is equal to the derivative of the free-energy with respect
to $n$ in $n=1$, this implies that the only invariants that enter in
the effective potential are the $J_k$. We would like to argue
that the same invariants are also the ones that govern fluctuations.

In the  basis (\ref{base}) the linear and quadratic 
invariants read:
\begin{eqnarray}
&&I_1=(n-1)\psi+\hat{\psi}\nonumber\\
&&J_2=(n-1)\psi^2 + 2 \psi \hat{\psi}+ \sum_{\mu}\chi_\mu^2+\hat{\psi}^2
\end{eqnarray}
so that the quadratic form writes: 
\begin{eqnarray}
\langle \phi | M | \phi \rangle = m_0 \left[ (n-1)\psi^2 + 2 \psi \hat{\psi}+ \sum_{\mu}\chi_\mu^2+\hat{\psi}^2 \right]
+\eta \left( (n-1)\psi+\hat{\psi} \right)^2. 
\end{eqnarray}
We now proceed with dimensional analysis, which as it is well known 
in general, is equivalent to the 
analysis of the leading singularities in perturbation theory.  

Imposing that for $n\to 1$ the terms $m_0 \psi\hat{\psi}$,
$m_0 \sum_\mu\chi_\mu^2$ and $\eta \hat{\psi}^2$ share the same superficial 
scaling
dimension, as $m_0 \to 0$ we find
\begin{eqnarray}
&& [\hat{\psi}] =2 +[\psi]\nonumber\\
&& [\chi_\mu]= 1  +[\psi].
\end{eqnarray}
where we have set $[m_0]=2$. Among the invariants of order $k$ the
ones of lower scaling dimension are these which contain the lower
power of $\hat{\psi}$ and $\chi_\mu$ for $n\to 1$. These are the terms
$J_k=\sum_\mu \psi_\mu^k$, which are the only ones that contain
$\psi^{k-1} \hat{\psi}$ and $\psi^{k-2} \sum_\mu \chi_\mu^2$. This is
enough to say that the spinodal lines, where the coefficient of $J_3$
is non-null, belong to the universality class of the $\phi^3$-RFIM
theory, (the spinodal of the RFIM).  At the critical point, by
definition the coefficient of $J_3$ in the effective action vanishes,
however in general, the coefficients of the other cubic invariants
$I_{3,1}$ and $I_{3,2}$ are non-zero. In order derive the RFIM, we
should argue that close to the upper critical dimension of the RFIM,
$D_c=6$,  these invariants have superficial scaling dimension higher
than the one of $J_4$.
 
The scaling dimension of $J_4$ is $[\hat{\psi}\psi^3]$, the ones of
$I_{3,2}$ and $I_{3,1}$ are respectively $[\hat{\psi}^2 \psi]$ and
$[\hat{\psi}^3]$. Since the dimension of $x$ is $-1$, in order to make
the action adimensional we need $[\psi]=D/2-2$.  We see that
$[{J_4}]=2 D-6$ while $[{I_{3,2}}]=3/2 D-2$ and $I_{3,1}=3/2D$. We
find therefore that close to the dimension 6, if the coefficient of
$J_3$ vanishes, the leading singular term becomes $J_4$ and the
critical point is in the RFIM class. Calling $g$ the coefficient of
$J_4$ in the Landau expansion of the free-energy, we can write
explicitly, close to the critical point and for $n\to 1$
\begin{eqnarray}
\label{PS}
  F[\psi]=&& \int dx \;\hat{\psi}(x)\left( -k \Delta \psi(x)
+m_0 \psi(x)+g \psi^3(x)+\eta \hat{\psi}(x)\right) \nonumber\\
&& +\frac 1 2\int dx\;\sum_\mu\left( k(\nabla\chi_\mu)^2+[m_0+
3g\psi(x)^2]\chi_\mu^2\right) 
\end{eqnarray}
It is well known that since there are $n-3\to - 2$ independent parameter
$\chi_\mu$ integration over them is equivalent to a Fermionic
determinant, and (\ref{PS}) is equivalent to the Parisi-Sourlas action
of the RFIM. Analogously to the case of the dynamical transition
\cite{FPRR11}, the fluctuations of the potential with respect to the
reference configurations can be effectively parametrized by a random
field term, with Gaussian statistics, uncorrelated from site to site.

\section{Correlation functions}
\label{sec:corr}
So far we have proceeded to a formal analysis of the replica soft
modes, we found that there are different components of the
fluctuations that have different scaling dimensions and we derived the
RFIM on the basis of a dimensional analysis. We anticipated in the
previous sections that the emergence of the RFIM comes from
fluctuations with respect to the choice of the reference
configuration. To substantiate this statement we study correlation
functions in our system relating them to corresponding function in the
RFIM.  We show that functions that are sensitive to thermal
fluctuations relate to thermal fluctuations of the RFIM, while
functions that are sensitive to the choice of the reference
configuration relate to functions sensitive to the choice of the
random field in the RFIM. 

 We define therefore two averages: we denote by angular
brackets $\langle\cdot\rangle$ the thermal average conditioned by the
choice of the reference configuration and can involve several
replicas, and by square brackets $[\cdot ]$ the average over the
choice of the reference configuration. Moreover, inside the averages,
index $0$ is assigned to the reference configuration, while indexes
$1,2,3,4$ refer to copies with different realization of the thermal
noise but subject to the attraction to the same reference
configuration.

We can write in principle seven distinct two-field (or four-body) correlation
functions, whose physical meaning is transparent:
\begin{eqnarray}
\label{eq:corr}
&& g_{0101}(x)=[\langle \phi_{01}(x)\phi_{01}(0)\rangle]; \;\;\;  
g_{0102}(x)=[\langle \phi_{01}(x) \phi_{02}(0)\rangle]=  
[\langle \phi_{01}(x)\rangle \langle \phi_{01}(0)\rangle]
\nonumber
\\
&&
g_{0112}(x) =[\langle \phi_{01}(x)\phi_{12}(0)\rangle];
\;\;\; g_{0123}(x)= [\langle 
\phi_{01}(x)\phi_{23}(0)
\rangle]\nonumber
\\
&& 
g_{1212}(x)=[\langle \phi_{12}(x)\phi_{12}(0)\rangle];\;\; 
g_{1213}(x)=[\langle \phi_{12}(x)\phi_{13}(0)\rangle]\nonumber
\\
&& 
g_{1234}(x)=[\langle \phi_{12}(x)\phi_{34}(0)\rangle]
\end{eqnarray}
To the order of the leading singularity, in which we can use the RFIM,
however, starting from the fields $\psi$, $\hat{\psi}$ and $\chi_\mu$
we can construct at most four independent correlations, namely
$\langle \psi(x)\psi(0)\rangle$, $\langle
\psi(x)\hat{\psi}(0)\rangle$, $\langle
\hat{\psi}(x)\hat{\psi}(0)\rangle$, and $\sum_\mu \langle
\chi_\mu(x)\chi_\mu(0)\rangle$, where we have taken into account the
condition $\sum_\mu \chi_\mu=0$.  This number is reduced to two by
replica symmetry, that implies that
\begin{eqnarray}
\label{iden}
&&  \sum_{a,b}^{1,n-1} \langle \phi_{0a}(x)\phi_{0b}(0)\rangle =(n-1)[g_{0101}-g_{0102}]+O((n-1)^2)
\nonumber
\\
&& \sum_{a,b,c,d}^{1,n-1}\langle \phi_{ab}(x)\phi_{cd}(0)\rangle =
(n-1)[-2g_{1212}+8g_{1213}-6g_{1234}]+O((n-1)^2)
\end{eqnarray}
The explicit expression in terms of $\psi$, $\hat{\psi}$ and
$\chi_\mu$, that we have computed with our Mathematica script, shows
that (\ref{iden}) are of order $n-1$ only if the following identities hold:
\begin{eqnarray}
&& \langle \hat{\psi}(x)\hat{\psi}(0)\rangle=0\nonumber\\
&& \sum_\mu \langle
\chi_\mu(x)\chi_\mu(0)\rangle=- \langle \psi(x)\hat{\psi}(0) +
\psi(0)\hat{\psi}(x) \rangle. 
\end{eqnarray}
These are known identities in the supersymmetric formalism for the
RFIM and related problems \cite{CGG99,CGPM03,CGP05} if we identify the
$\chi_\mu$ with Fermion fields.  It follows that the correlations
(\ref{eq:corr}) are linear combinations of $\langle
\psi(x)\hat{\psi}(0)\rangle$ and $\langle \psi(x)\psi(0)\rangle$,
which in the RFIM represent respectively the thermal fluctuations and
the sample to sample fluctuations correlation functions. The
coefficients of the combination depends on $\gamma= \frac{dq(p)}{dp}$
which is the only parameter of the system appearing in the
eigenvectors.

The correlations have connected and disconnected components with
respect to the angular brackets.  Correlations which are connected
measure thermal fluctuations. Disconnected correlations measure
instead fluctuations with respect to changes of the reference configuration.
Omitting the position indexes, we can write the connected combinations
$g_{0101}-g_{0102}$, $g_{0112}-g_{0123}$, $g_{1212}-g_{1213}$ and
$g_{1212}-g_{1234}$,
and the disconnected ones $g_{0102}$, $g_{0123}$, 
$g_{1213}$, and $g_{1234}$.
Exact expressions can be readily obtained with
Mathematica and read:
\begin{eqnarray}
\label{r1}
&& g_{0101}(x)-g_{0102}(x)=\frac{ 1 }{2-\gamma^2}\langle \psi(x)\hat{\psi}(0)+
\psi(0)\hat{\psi}(x)\rangle 
\nonumber
\\
&&
 g_{0112}(x)-g_{0123}(x)=\frac \gamma 2 ( g_{0101}(x)-g_{0102}(x) )
\nonumber
  \\
&& g_{1212}(x)-g_{1213}(x)=\frac{ \gamma^2}{ 4} ( g_{0101}(x)-g_{0102}(x) )
\nonumber
\\
&& g_{1212}(x)-g_{1234}(x)=\frac{ \gamma^2}{ 2} ( g_{0101}(x)-g_{0102}(x) )
\end{eqnarray}
The non connected components are
\begin{eqnarray}
\label{r2}
  && g_{0102}(x)=\frac{1}{2-\gamma^2}\langle \psi(x)\psi(0)\rangle
-\frac{4-\gamma^2}{2(2-\gamma^2)^2}\langle \psi(x)\hat{\psi}(0)+
\psi(0)\hat{\psi}(x)\rangle\nonumber
\\
&& g_{0123}(x)=\gamma \; g_{0102}(x)
\nonumber
  \\
&& g_{1213}(x)=\frac{\gamma^2}{2-\gamma^2}\langle \psi(x)\psi(0)\rangle
-\frac{\gamma^2(6-\gamma^2)}{4(2-\gamma^2)^2}\langle \psi(x)\hat{\psi}(0)+
\psi(0)\hat{\psi}(x)\rangle\nonumber
\\
&& g_{1234}(x)=\gamma^2\; g_{0102}(x)
\end{eqnarray}
As announced, we find that the connected correlations only contain 
$\langle \psi(x)\hat{\psi}(0)+
\psi(0)\hat{\psi}(x)\rangle$ that in the RFIM represents the thermal
correlation function for fixed random field while the disconnected
correlations also contain $\langle \psi(x)\psi(0)\rangle$ which is the 
correlation sensitive to random field changes in the RFIM. 

The relations (\ref{r1},\ref{r2}) between the different overlap
correlation functions are an important consequence of our analysis,
they are valid at all orders of perturbation theory and can be tested
in numerical simulations. The computation of all correlations
(\ref{eq:corr}) requires to simulate a maximum of four independent
replicas besides the reference one.  The parameter
$\gamma$ can also in principle be measured in simulations by looking
on the dependence on $p$ of the overlap $q$ between two distinct
replicas that have overlap $p$ with the reference. 

The relations (\ref{r1},\ref{r2}) express the fact that to the leading
order the overlap of the system with the reference configuration and
the self-overlaps are strongly correlated and the fluctuations verify
$\phi_{12}(x)=\delta q_{12}(x)\sim \gamma \phi_{01}(x)=\delta
q_{01}(x)$. The second of (\ref{r1}) for example can be derived
observing that both the LHS and the RHS combinations can be expressed
as derivative of local overlap averages with respect to space
dependent coupling $\epsilon(x)$,
\begin{eqnarray}
  \label{eq:ww}
&&g_{0101}(x)-g_{0102}(x)=\frac 1 T \frac{\delta [\langle q_{01}(x) \rangle]}{\delta \epsilon(0)};\;\;\;\; 
g_{0112}(x)-g_{0123}(x)=\frac {1}{2 T} \frac{\delta [\langle q_{12}(x) \rangle]}{\delta \epsilon(0)}  
\end{eqnarray}
and using the chain rule for the derivative
\begin{eqnarray}
  \label{eq:chain}
\frac{\delta [\langle q_{12}(x) \rangle]}{\delta \epsilon(0)}=
\frac{dq(p)}{dp}\frac{\delta [\langle q_{01}(x) \rangle]}{\delta \epsilon(0)}.  
\end{eqnarray}
It can be noted however that when integrated over space the second of
(\ref{r1}) holds exactly even beyond the level of leading singularity
described by the RFIM and can be used to measure $\gamma$.

Finally, we remark that as it is well
known in the theory of the RFIM, to the one-loop order of Gaussian
fluctuations, the non-connected components are more
singular than the connected ones. In momentum space, connected
correlations behave as $(k^2+m_0)^{-1}$, while disconnected ones
behave as $(k^2+m_0)^{-2}$.

\section{Summary and Conclusions} 
In this paper we have analyzed the universality class of critical
points terminating first order transition lines in glassy systems in
presence of constraints. After the analysis of the case of a symmetric
coupling between two replicas, for which the Ising universality class
is found, we have considered the case in which a coupling with a
quenched reference configuration is present. We extend in this way the
analysis of the equal temperature pinned particle critical point
presented in \cite{FPR13}. This includes the quenched potential
construction and the pinned particle construction for $T_{ref}<T$.  A
full analysis of the soft modes within Replica Field Theory at all
order of perturbation theory leads to the universality class of the
Random Field Ising Model. The effective random field appearing in the
final description parametrizes the randomness in the reference
configuration. We have analyzed the various four-body correlation
functions that appear in the theory and found that there are only two
independent combinations that become dominant close to the critical
point. The existence of the critical point, its universality class and
the relation between correlation functions constitute important
predictions of the Thermodynamic theory of glasses based on Mean Field
Theory, and now, on its loop expansion. We hope that in a next future
they can be tested in numerical simulations of realistic glass forming
liquid models.  The critical point and line of continuous glass
transition present in the particle pinning problem for $T_{ref}>T$ are
excluded by the present analysis.  In that case the replicon modes are
critical and their interaction with the longitudinal and anomalous
modes that we have seen to give rise to the RFIM should be included.
This is a
fascinating research project that we leave for the future.
\begin{center}
*
\end{center}

After completion of our work we came to know that G. Biroli,
C. Cammarota, G. Tarjus and M. Tarzia have also considered the problem
of quenched critical points with a similar approach (arXiv:1309.3194).

\begin{center}
{\bf  Acknowledgments}
\end{center}
We thank L. Berthier, F. Ricci-Tersenghi, T. Rizzo, S. Sastry and
P. Urbani, for useful discussions and exchanges. We also thank our
JSTAT anonymous referee for insightful remarks and constructive
comments. SF thanks the Dipartimento di Fisica Universit\`a di Roma
``La Sapienza'' for hospitality.  The European Research Council has
provided financial support through ERC Grant 247328.

\section{Appendix} 
The algebra of multiplication of the four replica index mass matrix
$M[a,b;c,d]$ by a two replica index vectors $v_{ab}$, which is
necessary to perform the explicit computation of the eigenvalues, can
be implemented in Mathematica. In fact, given the structure of $M$ and
its eigenvalues, one just needs to be able to perform sums over
replica indexes of constants and delta functions. A detailed script
performing this task was published in \cite{FPR13}. The one we use
here is an adaptation of that one. The eigenvalues of the mass matrix
can be then explicitly found. At the end, this amounts to solving
second order algebraic equations.  We present the result of the
computation to the order $(n-1)$.  The longitudinal eigenvalues read:
\begin{eqnarray}
&&\lambda_{LO}^{\pm}=
\frac{1}{4} \left(4 m_1-2 m_2+\mu _2\pm \sqrt{\left(\mu _2+2 m_2\right){}^2-8 \nu
   _2^2}\right)+
\\
&&+\frac{1}{4} (n-1) \left(2 m_2-2 m_3+\mu _2+4 \mu
   _3 \pm \frac{4 \nu _2 \left(\nu _2-4 \nu _3\right)-\left(\mu _2+2
   m_2\right) \left(-\mu _2-4 \mu _3+2 m_2-2 m_3\right)}{\sqrt{\left(\mu
   _2+2 m_2\right){}^2-8 \nu _2^2}}\right)
\nonumber
\end{eqnarray}
The zero mode corresponds to the positive determination of the square
root $\lambda_{LO}=\lambda_{LO}^+$. This can be argued be the
condition that in the $S_n$ symmetric limit $\mu_2=\nu_2=m_2$ and
$\mu_3=\nu_3=m_3$ the leading order $m_0$ become degenerate with the
replicon eigenvalue $\lambda_{RE}=m_1$.  In this way we find:
\begin{eqnarray}
&&  m_0=\frac{1}{4} \left(4 m_1-2 m_2+\mu _2+ \sqrt{\left(\mu _2+2 m_2\right){}^2-8 \nu
   _2^2}\right)
\\
&&\eta_{LO}=\frac{1}{4} \left(2 m_2-2 m_3+\mu _2+4 \mu
   _3 + \frac{4 \nu _2 \left(\nu _2-4 \nu _3\right)-\left(\mu _2+2
   m_2\right) \left(-\mu _2-4 \mu _3+2 m_2-2 m_3\right)}{\sqrt{\left(\mu
   _2+2 m_2\right){}^2-8 \nu _2^2}}\right). 
\end{eqnarray}
We notice that for $n\to 1 $ the longitudinal eigenvalue becomes
degenerate with the replicon if $\nu_2=\pm \sqrt{|m_2 \mu_2 |}$. This
condition is met at the critical point of junction of the
discontinuous and continuous glass transition of the pinned particle
problem for $T_{ref}>T$, and actually on the whole line of continuous
glass transition.

Analogously one can compute the anomalous eigenvalues that 
turn out to be: 
\begin{eqnarray}
  \lambda_{AN}^\pm=&&
\frac{1}{4} \left(4 m_1-2 m_2+\mu _2\pm \sqrt{\left(\mu _2+2 m_2\right){}^2-8 \nu
   _2^2}\right)+\\
&&\pm \frac{(n-1) \nu _2^2 \left(\mu _2 \left(\mu
   _2\pm\sqrt{\left(\mu _2+2 m_2\right){}^2-8 \nu _2^2}+2 m_2\right)-4 \nu
   _2^2\right)}{\sqrt{\left(\mu _2+2 m_2\right){}^2-8 \nu _2^2}
   \left(\mu _2\pm\sqrt{\left(\mu _2+2 m_2\right){}^2-8 \nu _2^2}+2
   m_2\right){}^2}\nonumber
\end{eqnarray}
with $\lambda_{AN}=\lambda_{AN}^+$ and 
\begin{eqnarray}
  \eta_{AN}=\frac{\nu _2^2 \left(\mu _2 \left(\mu
   _2+\sqrt{\left(\mu _2+2 m_2\right){}^2-8 \nu _2^2}+2 m_2\right)-4 \nu
   _2^2\right)}{\sqrt{\left(\mu _2+2 m_2\right){}^2-8 \nu _2^2}
   \left(\mu _2+\sqrt{\left(\mu _2+2 m_2\right){}^2-8 \nu _2^2}+2
   m_2\right){}^2}.
\end{eqnarray}
The expression of $\eta=\eta_{LO}-\eta_{AN}$ is not particularly
illuminating.  The consistency of the approach requires that it should
remain positive at the critical point.

\bibliography{BPD}{}

%Merlin.mbs v4.21 2009-07-09.
\begin{thebibliography}{10}%
\makeatletter
\providecommand \@ifxundefined [1]{%
 \ifx #1\undefined \expandafter \@firstoftwo
 \else \expandafter \@secondoftwo
\fi
}%
\providecommand \@ifnum [1]{%
 \ifnum #1\expandafter \@firstoftwo
 \else \expandafter \@secondoftwo
\fi
}%
\providecommand \enquote [1]{``#1''}%
\providecommand \bibnamefont  [1]{#1}%
\providecommand \bibfnamefont [1]{#1}%
\providecommand \citenamefont [1]{#1}%
\providecommand\href[0]{\@sanitize\@href}%
\providecommand\@href[1]{\endgroup\@@startlink{#1}\endgroup\@@href}%
\providecommand\@@href[1]{#1\@@endlink}%
\providecommand \@sanitize [0]{\begingroup\catcode`\&12\catcode`\#12\relax}%
\@ifxundefined \pdfoutput {\@firstoftwo}{%
 \@ifnum{\z@=\pdfoutput}{\@firstoftwo}{\@secondoftwo}%
}{%
 \providecommand\@@startlink[1]{\leavevmode\special{html:<a href="#1">}}%
 \providecommand\@@endlink[0]{\special{html:</a>}}%
}{%
 \providecommand\@@startlink[1]{%
  \leavevmode
  \pdfstartlink
   attr{/Border[0 0 1 ]/H/I/C[0 1 1]}%
   user{/Subtype/Link/A<</Type/Action/S/URI/URI(#1)>>}%
  \relax
 }%
 \providecommand\@@endlink[0]{\pdfendlink}%
}%
\providecommand \url  [0]{\begingroup\@sanitize \@url }%
\providecommand \@url [1]{\endgroup\@href {#1}{\urlprefix}}%
\providecommand \urlprefix [0]{URL }%
\providecommand \Eprint[0]{\href }%
\@ifxundefined \urlstyle {%
  \providecommand \doi [1]{doi:\discretionary{}{}{}#1}%
}{%
  \providecommand \doi [0]{doi:\discretionary{}{}{}\begingroup
  \urlstyle{rm}\Url }%
}%
\providecommand \doibase [0]{http://dx.doi.org/}%
\providecommand \Doi[1]{\href{\doibase#1}}%
\providecommand \bibAnnote [3]{%
  \BibitemShut{#1}%
  \begin{quotation}\noindent
    \textsc{Key:}\ #2\\\textsc{Annotation:}\ #3%
  \end{quotation}%
}%
\providecommand \bibAnnoteFile [2]{%
  \IfFileExists{#2}{\bibAnnote {#1} {#2} {\input{#2}}}{}%
}%
\providecommand \typeout [0]{\immediate \write \m@ne }%
\providecommand \selectlanguage [0]{\@gobble}%
\providecommand \bibinfo [0]{\@secondoftwo}%
\providecommand \bibfield [0]{\@secondoftwo}%
\providecommand \translation [1]{[#1]}%
\providecommand \BibitemOpen[0]{}%
\providecommand \bibitemStop [0]{}%
\providecommand \bibitemNoStop [0]{.\EOS\space}%
\providecommand \EOS [0]{\spacefactor3000\relax}%
\providecommand \BibitemShut [1]{\csname bibitem#1\endcsname}%
%</preamble>
\bibitem{Ca09}%
  \BibitemOpen
  \bibfield{author}{%
  \bibinfo {author} {\bibfnamefont{A.}~\bibnamefont{Cavagna}},\ }%
  \bibfield{journal}{%
  \bibinfo {journal} {Physics Reports}\ }%
  \textbf{\bibinfo {volume} {476}},\ \bibinfo {pages} {51} (\bibinfo {year}
  {2009})%
  \bibAnnoteFile{NoStop}{Ca09}%
\bibitem{WL12}%
  \BibitemOpen
  \bibfield{author}{%
  \bibinfo {author} {\bibfnamefont{P.~G.}\ \bibnamefont{Wolynes}}\ and\
  \bibinfo {author} {\bibfnamefont{V.}~\bibnamefont{Lubchenko}},\ }%
  \emph{\bibinfo {title} {Structural Glasses and Supercooled Liquids: Theory,
  Experiment, and Applications}}\ (\bibinfo {publisher} {Wiley. com},\ \bibinfo
  {year} {2012})%
  \bibAnnoteFile{NoStop}{WL12}%
\bibitem{KTW89}%
  \BibitemOpen
  \bibfield{author}{%
  \bibinfo {author} {\bibfnamefont{T.~R.}\ \bibnamefont{Kirkpatrick}}, \bibinfo
  {author} {\bibfnamefont{D.}~\bibnamefont{Thirumalai}},\ and\ \bibinfo
  {author} {\bibfnamefont{P.~G.}\ \bibnamefont{Wolynes}},\ }%
  \bibfield{journal}{%
  \Doi{10.1103/PhysRevA.40.1045}{\bibinfo {journal} {Phys. Rev. A}}\ }%
  \textbf{\bibinfo {volume} {40}},\ \bibinfo {pages} {1045} (\bibinfo {month}
  {Jul}\ \bibinfo {year} {1989})%
  \bibAnnoteFile{NoStop}{KTW89}%
\bibitem{FPV92}%
  \BibitemOpen
  \bibfield{author}{%
  \bibinfo {author} {\bibfnamefont{S.}~\bibnamefont{Franz}}, \bibinfo {author}
  {\bibfnamefont{G.}~\bibnamefont{Parisi}},\ and\ \bibinfo {author}
  {\bibfnamefont{M.}~\bibnamefont{Virasoro}},\ }%
  \bibfield{journal}{%
  \bibinfo {journal} {Journal de Physique I}\ }%
  \textbf{\bibinfo {volume} {2}},\ \bibinfo {pages} {1869} (\bibinfo {year}
  {1992})%
  \bibAnnoteFile{NoStop}{FPV92}%
\bibitem{FP95}%
  \BibitemOpen
  \bibfield{author}{%
  \bibinfo {author} {\bibfnamefont{S.}~\bibnamefont{Franz}}\ and\ \bibinfo
  {author} {\bibfnamefont{G.}~\bibnamefont{Parisi}},\ }%
  \bibfield{journal}{%
  \bibinfo {journal} {Journal de Physique I}\ }%
  \textbf{\bibinfo {volume} {5}},\ \bibinfo {pages} {1401} (\bibinfo {year}
  {1995})%
  \bibAnnoteFile{NoStop}{FP95}%
\bibitem{Kim03}%
  \BibitemOpen
  \bibfield{author}{%
  \bibinfo {author} {\bibfnamefont{K.}~\bibnamefont{Kim}},\ }%
  \bibfield{journal}{%
  \bibinfo {journal} {EPL (Europhysics Letters)}\ }%
  \textbf{\bibinfo {volume} {61}},\ \bibinfo {pages} {790} (\bibinfo {year}
  {2003})%
  \bibAnnoteFile{NoStop}{Kim03}%
\bibitem{BK12}%
  \BibitemOpen
  \bibfield{author}{%
  \bibinfo {author} {\bibfnamefont{L.}~\bibnamefont{Berthier}}\ and\ \bibinfo
  {author} {\bibfnamefont{W.}~\bibnamefont{Kob}},\ }%
  \bibfield{journal}{%
  \Doi{10.1103/PhysRevE.85.011102}{\bibinfo {journal} {Phys. Rev. E}}\ }%
  \textbf{\bibinfo {volume} {85}},\ \bibinfo {pages} {011102} (\bibinfo {month}
  {Jan}\ \bibinfo {year} {2012}),\
  \url{http://link.aps.org/doi/10.1103/PhysRevE.85.011102}%
  \bibAnnoteFile{NoStop}{BK12}%
\bibitem{KB13}%
  \BibitemOpen
  \bibfield{author}{%
  \bibinfo {author} {\bibfnamefont{W.}~\bibnamefont{Kob}}\ and\ \bibinfo
  {author} {\bibfnamefont{L.}~\bibnamefont{Berthier}},\ }%
  \bibfield{journal}{%
  \bibinfo {journal} {Physical Review Letters}\ }%
  \textbf{\bibinfo {volume} {110}},\ \bibinfo {pages} {245702} (\bibinfo {year}
  {2013})%
  \bibAnnoteFile{NoStop}{KB13}%
\bibitem{BC12}%
  \BibitemOpen
  \bibfield{author}{%
  \bibinfo {author} {\bibfnamefont{C.}~\bibnamefont{Cammarota}}\ and\ \bibinfo
  {author} {\bibfnamefont{G.}~\bibnamefont{Biroli}},\ }%
  \bibfield{journal}{%
  \bibinfo {journal} {Proceedings of the National Academy of Sciences}\ }%
  \textbf{\bibinfo {volume} {109}},\ \bibinfo {pages} {8850} (\bibinfo {year}
  {2012})%
  \bibAnnoteFile{NoStop}{BC12}%
\bibitem{KP13}%
  \BibitemOpen
  \bibfield{author}{%
  \bibinfo {author} {\bibfnamefont{S.}~\bibnamefont{Karmakar}}\ and\ \bibinfo
  {author} {\bibfnamefont{G.}~\bibnamefont{Parisi}},\ }%
  \bibfield{journal}{%
  \bibinfo {journal} {Proceedings of the National Academy of Sciences}\ }%
  \textbf{\bibinfo {volume} {110}},\ \bibinfo {pages} {2752} (\bibinfo {year}
  {2013})%
  \bibAnnoteFile{NoStop}{KP13}%
\bibitem{Go08}%
  \BibitemOpen
  \bibfield{author}{%
  \bibinfo {author} {\bibfnamefont{W.}~\bibnamefont{G{\"o}tze}},\ }%
  \emph{\bibinfo {title} {Complex Dynamics of Glass-Forming Liquids: A
  Mode-Coupling Theory: A Mode-Coupling Theory}},\ Vol.\ \bibinfo {volume}
  {143}\ (\bibinfo {publisher} {Oxford University Press},\ \bibinfo {year}
  {2008})%
  \bibAnnoteFile{NoStop}{Go08}%
\bibitem{FP00}%
  \BibitemOpen
  \bibfield{author}{%
  \bibinfo {author} {\bibfnamefont{S.}~\bibnamefont{Franz}}\ and\ \bibinfo
  {author} {\bibfnamefont{G.}~\bibnamefont{Parisi}},\ }%
  \bibfield{journal}{%
  \bibinfo {journal} {Journal of Physics: Condensed Matter}\ }%
  \textbf{\bibinfo {volume} {12}},\ \bibinfo {pages} {6335} (\bibinfo {year}
  {2000}),\ \url{http://stacks.iop.org/0953-8984/12/i=29/a=305}%
  \bibAnnoteFile{NoStop}{FP00}%
\bibitem{DFGP02}%
  \BibitemOpen
  \bibfield{author}{%
  \bibinfo {author} {\bibfnamefont{C.}~\bibnamefont{Donati}}, \bibinfo {author}
  {\bibfnamefont{S.}~\bibnamefont{Franz}}, \bibinfo {author}
  {\bibfnamefont{S.}~\bibnamefont{Glotzer}},\ and\ \bibinfo {author}
  {\bibfnamefont{G.}~\bibnamefont{Parisi}},\ }%
  \bibfield{journal}{%
  \bibinfo {journal} {Journal of non-crystalline solids}\ }%
  \textbf{\bibinfo {volume} {307}},\ \bibinfo {pages} {215} (\bibinfo {year}
  {2002})%
  \bibAnnoteFile{NoStop}{DFGP02}%
\bibitem{FPRR11}%
  \BibitemOpen
  \bibfield{author}{%
  \bibinfo {author} {\bibfnamefont{S.}~\bibnamefont{Franz}}, \bibinfo {author}
  {\bibfnamefont{G.}~\bibnamefont{Parisi}}, \bibinfo {author}
  {\bibfnamefont{F.}~\bibnamefont{Ricci-Tersenghi}},\ and\ \bibinfo {author}
  {\bibfnamefont{T.}~\bibnamefont{Rizzo}},\ }%
  \bibfield{journal}{%
  \bibinfo {journal} {The European Physical Journal E: Soft Matter and
  Biological Physics}\ }%
  \textbf{\bibinfo {volume} {34}},\ \bibinfo {pages} {1} (\bibinfo {year}
  {2011})%
  \bibAnnoteFile{NoStop}{FPRR11}%
\bibitem{FJPUZ13-1}%
  \BibitemOpen
  \bibfield{author}{%
  \bibinfo {author} {\bibfnamefont{S.}~\bibnamefont{Franz}}, \bibinfo {author}
  {\bibfnamefont{H.}~\bibnamefont{Jacquin}}, \bibinfo {author}
  {\bibfnamefont{G.}~\bibnamefont{Parisi}}, \bibinfo {author}
  {\bibfnamefont{P.}~\bibnamefont{Urbani}},\ and\ \bibinfo {author}
  {\bibfnamefont{F.}~\bibnamefont{Zamponi}},\ }%
  \bibfield{journal}{%
  \bibinfo {journal} {Proceedings of the National Academy of Sciences of the
  United States of America}\ }%
  \textbf{\bibinfo {volume} {110}},\ \bibinfo {pages} {11211} (\bibinfo {year}
  {2013})%
  \bibAnnoteFile{NoStop}{FJPUZ13-1}%
\bibitem{FJPUZ13-2}%
  \BibitemOpen
  \bibfield{author}{%
  \bibinfo {author} {\bibfnamefont{S.}~\bibnamefont{Franz}}, \bibinfo {author}
  {\bibfnamefont{H.}~\bibnamefont{Jacquin}}, \bibinfo {author}
  {\bibfnamefont{G.}~\bibnamefont{Parisi}}, \bibinfo {author}
  {\bibfnamefont{P.}~\bibnamefont{Urbani}},\ and\ \bibinfo {author}
  {\bibfnamefont{F.}~\bibnamefont{Zamponi}},\ }%
  \bibfield{journal}{%
  \bibinfo {journal} {The Journal of chemical physics}\ }%
  \textbf{\bibinfo {volume} {138}},\ \bibinfo {pages} {12A540} (\bibinfo {year}
  {2013})%
  \bibAnnoteFile{NoStop}{FJPUZ13-2}%
\bibitem{FKZ12}%
  \BibitemOpen
  \bibfield{author}{%
  \bibinfo {author} {\bibfnamefont{L.}~\bibnamefont{Foini}}, \bibinfo {author}
  {\bibfnamefont{F.}~\bibnamefont{Krzakala}},\ and\ \bibinfo {author}
  {\bibfnamefont{F.}~\bibnamefont{Zamponi}},\ }%
  \bibfield{journal}{%
  \bibinfo {journal} {Journal of Statistical Mechanics: Theory and Experiment}\
  }%
  \textbf{\bibinfo {volume} {2012}},\ \bibinfo {pages} {P06013} (\bibinfo
  {year} {2012})%
  \bibAnnoteFile{NoStop}{FKZ12}%
\bibitem{FS13}%
  \BibitemOpen
  \bibfield{author}{%
  \bibinfo {author} {\bibfnamefont{S.}~\bibnamefont{Franz}}\ and\ \bibinfo
  {author} {\bibfnamefont{M.}~\bibnamefont{Sellitto}},\ }%
  \bibfield{journal}{%
  \bibinfo {journal} {Journal of Statistical Mechanics: Theory and Experiment}\
  }%
  \textbf{\bibinfo {volume} {2013}},\ \bibinfo {pages} {P02025} (\bibinfo
  {year} {2013})%
  \bibAnnoteFile{NoStop}{FS13}%
\bibitem{FP97}%
  \BibitemOpen
  \bibfield{author}{%
  \bibinfo {author} {\bibfnamefont{S.}~\bibnamefont{Franz}}\ and\ \bibinfo
  {author} {\bibfnamefont{G.}~\bibnamefont{Parisi}},\ }%
  \bibfield{journal}{%
  \Doi{10.1103/PhysRevLett.79.2486}{\bibinfo {journal} {Phys. Rev. Lett.}}\ }%
  \textbf{\bibinfo {volume} {79}},\ \bibinfo {pages} {2486} (\bibinfo {month}
  {Sep}\ \bibinfo {year} {1997})%
  \bibAnnoteFile{NoStop}{FP97}%
\bibitem{FP98}%
  \BibitemOpen
  \bibfield{author}{%
  \bibinfo {author} {\bibfnamefont{S.}~\bibnamefont{Franz}}\ and\ \bibinfo
  {author} {\bibfnamefont{G.}~\bibnamefont{Parisi}},\ }%
  \bibfield{journal}{%
  \bibinfo {journal} {Physica A: Statistical Mechanics and its Applications}\
  }%
  \textbf{\bibinfo {volume} {261}},\ \bibinfo {pages} {317} (\bibinfo {year}
  {1998})%
  \bibAnnoteFile{NoStop}{FP98}%
\bibitem{CaBi13}%
  \BibitemOpen
  \bibfield{author}{%
  \bibinfo {author} {\bibfnamefont{C.}~\bibnamefont{Cammarota}}\ and\ \bibinfo
  {author} {\bibfnamefont{G.}~\bibnamefont{Biroli}},\ }%
  \bibfield{journal}{%
  \bibinfo {journal} {The Journal of chemical physics}\ }%
  \textbf{\bibinfo {volume} {138}},\ \bibinfo {pages} {12A547} (\bibinfo {year}
  {2013})%
  \bibAnnoteFile{NoStop}{CaBi13}%
\bibitem{Ca13}%
  \BibitemOpen
  \bibfield{author}{%
  \bibinfo {author} {\bibfnamefont{C.}~\bibnamefont{Cammarota}},\ }%
  \bibfield{journal}{%
  \bibinfo {journal} {EPL (Europhysics Letters)}\ }%
  \textbf{\bibinfo {volume} {101}},\ \bibinfo {pages} {56001} (\bibinfo {year}
  {2013})%
  \bibAnnoteFile{NoStop}{Ca13}%
\bibitem{GC02}%
  \BibitemOpen
  \bibfield{author}{%
  \bibinfo {author} {\bibfnamefont{J.~P.}\ \bibnamefont{Garrahan}}\ and\
  \bibinfo {author} {\bibfnamefont{D.}~\bibnamefont{Chandler}},\ }%
  \bibfield{journal}{%
  \bibinfo {journal} {Physical review letters}\ }%
  \textbf{\bibinfo {volume} {89}},\ \bibinfo {pages} {035704} (\bibinfo {year}
  {2002})%
  \bibAnnoteFile{NoStop}{GC02}%
\bibitem{GST11}%
  \BibitemOpen
  \bibfield{author}{%
  \bibinfo {author} {\bibfnamefont{J.~P.}\ \bibnamefont{Garrahan}}, \bibinfo
  {author} {\bibfnamefont{P.}~\bibnamefont{Sollich}},\ and\ \bibinfo {author}
  {\bibfnamefont{C.}~\bibnamefont{Toninelli}},\ }%
  \bibfield{journal}{%
  \bibinfo {journal} {Dynamical Heterogeneities in Glasses, Colloids, and
  Granular Media}\ }%
  \textbf{\bibinfo {volume} {150}},\ \bibinfo {pages} {341} (\bibinfo {year}
  {2011})%
  \bibAnnoteFile{NoStop}{GST11}%
\bibitem{JB12}%
  \BibitemOpen
  \bibfield{author}{%
  \bibinfo {author} {\bibfnamefont{R.~L.}\ \bibnamefont{Jack}}\ and\ \bibinfo
  {author} {\bibfnamefont{L.}~\bibnamefont{Berthier}},\ }%
  \bibfield{journal}{%
  \bibinfo {journal} {Physical Review E}\ }%
  \textbf{\bibinfo {volume} {85}},\ \bibinfo {pages} {021120} (\bibinfo {year}
  {2012})%
  \bibAnnoteFile{NoStop}{JB12}%
\bibitem{CCGGPV10}%
  \BibitemOpen
  \bibfield{author}{%
  \bibinfo {author} {\bibfnamefont{C.}~\bibnamefont{Cammarota}}, \bibinfo
  {author} {\bibfnamefont{A.}~\bibnamefont{Cavagna}}, \bibinfo {author}
  {\bibfnamefont{I.}~\bibnamefont{Giardina}}, \bibinfo {author}
  {\bibfnamefont{G.}~\bibnamefont{Gradenigo}}, \bibinfo {author}
  {\bibfnamefont{T.}~\bibnamefont{Grigera}}, \bibinfo {author}
  {\bibfnamefont{G.}~\bibnamefont{Parisi}},\ and\ \bibinfo {author}
  {\bibfnamefont{P.}~\bibnamefont{Verrocchio}},\ }%
  \bibfield{journal}{%
  \bibinfo {journal} {Physical review letters}\ }%
  \textbf{\bibinfo {volume} {105}},\ \bibinfo {pages} {055703} (\bibinfo {year}
  {2010})%
  \bibAnnoteFile{NoStop}{CCGGPV10}%
\bibitem{Berthier}%
  \BibitemOpen
  \bibfield{author}{%
  \bibinfo {author} {\bibfnamefont{L.}~\bibnamefont{Berthier}},\ }%
  \bibfield{journal}{%
  \bibinfo {journal} {Phys. Rev. E}\ }%
  \textbf{\bibinfo {volume} {88}},\ \bibinfo {pages} {022313} (\bibinfo {year}
  {2013})%
  \bibAnnoteFile{NoStop}{Berthier}%
\bibitem{Note1}%
  \BibitemOpen
  \bibinfo {note} {G. Parisi and B. Seoane, In preparation}%
  \bibAnnoteFile{NoStop}{Note1}%
\bibitem{BY91}%
  \BibitemOpen
  \bibfield{author}{%
  \bibinfo {author} {\bibfnamefont{D.}~\bibnamefont{Belanger}}\ and\ \bibinfo
  {author} {\bibfnamefont{A.}~\bibnamefont{Young}},\ }%
  \bibfield{journal}{%
  \bibinfo {journal} {Journal of magnetism and magnetic materials}\ }%
  \textbf{\bibinfo {volume} {100}},\ \bibinfo {pages} {272} (\bibinfo {year}
  {1991})%
  \bibAnnoteFile{NoStop}{BY91}%
\bibitem{N98}%
  \BibitemOpen
  \bibfield{author}{%
  \bibinfo {author} {\bibfnamefont{T.}~\bibnamefont{Nattermann}},\ }%
  \bibfield{journal}{%
  \bibinfo {journal} {{\it in} P. Young (ed.) {\it Spin Glasses and Random
  Fields } World Scientific Singapore \;}}%
   (\bibinfo {year} {1998})%
  \bibAnnoteFile{NoStop}{N98}%
\bibitem{FPR13}%
  \BibitemOpen
  \bibfield{author}{%
  \bibinfo {author} {\bibfnamefont{S.}~\bibnamefont{Franz}}, \bibinfo {author}
  {\bibfnamefont{G.}~\bibnamefont{Parisi}},\ and\ \bibinfo {author}
  {\bibfnamefont{F.}~\bibnamefont{Ricci-Tersenghi}},\ }%
  \bibfield{journal}{%
  \bibinfo {journal} {Journal of Statistical Mechanics: Theory and Experiment}\
  }%
  \textbf{\bibinfo {volume} {2013}},\ \bibinfo {pages} {L02001} (\bibinfo
  {year} {2013})%
  \bibAnnoteFile{NoStop}{FPR13}%
\bibitem{PS82}%
  \BibitemOpen
  \bibfield{author}{%
  \bibinfo {author} {\bibfnamefont{G.}~\bibnamefont{Parisi}}\ and\ \bibinfo
  {author} {\bibfnamefont{N.}~\bibnamefont{Sourlas}},\ }%
  \bibfield{journal}{%
  \bibinfo {journal} {Nuclear Physics B}\ }%
  \textbf{\bibinfo {volume} {206}},\ \bibinfo {pages} {321} (\bibinfo {year}
  {1982})%
  \bibAnnoteFile{NoStop}{PS82}%
\bibitem{Me99}%
  \BibitemOpen
  \bibfield{author}{%
  \bibinfo {author} {\bibfnamefont{M.}~\bibnamefont{M{\'e}zard}},\ }%
  \bibfield{journal}{%
  \bibinfo {journal} {Physica A: Statistical Mechanics and its Applications}\
  }%
  \textbf{\bibinfo {volume} {265}},\ \bibinfo {pages} {352} (\bibinfo {year}
  {1999})%
  \bibAnnoteFile{NoStop}{Me99}%
\bibitem{CFP99}%
  \BibitemOpen
  \bibfield{author}{%
  \bibinfo {author} {\bibfnamefont{M.}~\bibnamefont{Cardenas}}, \bibinfo
  {author} {\bibfnamefont{S.}~\bibnamefont{Franz}},\ and\ \bibinfo {author}
  {\bibfnamefont{G.}~\bibnamefont{Parisi}},\ }%
  \bibfield{journal}{%
  \bibinfo {journal} {The Journal of chemical physics}\ }%
  \textbf{\bibinfo {volume} {110}},\ \bibinfo {pages} {1726} (\bibinfo {year}
  {1999})%
  \bibAnnoteFile{NoStop}{CFP99}%
\bibitem{CS92}%
  \BibitemOpen
  \bibfield{author}{%
  \bibinfo {author} {\bibfnamefont{A.}~\bibnamefont{Crisanti}}\ and\ \bibinfo
  {author} {\bibfnamefont{H.-J.}\ \bibnamefont{Sommers}},\ }%
  \bibfield{journal}{%
  \bibinfo {journal} {Zeitschrift f{\"u}r Physik B Condensed Matter}\ }%
  \textbf{\bibinfo {volume} {87}},\ \bibinfo {pages} {341} (\bibinfo {year}
  {1992})%
  \bibAnnoteFile{NoStop}{CS92}%
\bibitem{KrZd10}%
  \BibitemOpen
  \bibfield{author}{%
  \bibinfo {author} {\bibfnamefont{F.}~\bibnamefont{Krzakala}}\ and\ \bibinfo
  {author} {\bibfnamefont{L.}~\bibnamefont{Zdeborov{\'a}}},\ }%
  \bibfield{journal}{%
  \bibinfo {journal} {EPL (Europhysics Letters)}\ }%
  \textbf{\bibinfo {volume} {90}},\ \bibinfo {pages} {66002} (\bibinfo {year}
  {2010})%
  \bibAnnoteFile{NoStop}{KrZd10}%
\bibitem{CFP97}%
  \BibitemOpen
  \bibfield{author}{%
  \bibinfo {author} {\bibfnamefont{M.}~\bibnamefont{Cardenas}}, \bibinfo
  {author} {\bibfnamefont{S.}~\bibnamefont{Franz}},\ and\ \bibinfo {author}
  {\bibfnamefont{G.}~\bibnamefont{Parisi}},\ }%
  \bibfield{journal}{%
  \bibinfo {journal} {arXiv preprint cond-mat/9712099}}%
   (\bibinfo {year} {1997})%
  \bibAnnoteFile{NoStop}{CFP97}%
\bibitem{FS11}%
  \BibitemOpen
  \bibfield{author}{%
  \bibinfo {author} {\bibfnamefont{S.}~\bibnamefont{Franz}}, \bibinfo {author}
  {\bibfnamefont{G.}~\bibnamefont{Semerjian}}, \emph{et~al.},\ }%
  \bibfield{journal}{%
  \bibinfo {journal} {Dynamical Heterogeneities in Glasses, Colloids, and
  Granular Media}\ }%
  \textbf{\bibinfo {volume} {407}} (\bibinfo {year} {2011})%
  \bibAnnoteFile{NoStop}{FS11}%
\bibitem{Kr10}%
  \BibitemOpen
  \bibfield{author}{%
  \bibinfo {author} {\bibfnamefont{V.}~\bibnamefont{Krakoviack}},\ }%
  \bibfield{journal}{%
  \bibinfo {journal} {Physical Review E}\ }%
  \textbf{\bibinfo {volume} {82}},\ \bibinfo {pages} {061501} (\bibinfo {year}
  {2010})%
  \bibAnnoteFile{NoStop}{Kr10}%
\bibitem{Note2}%
  \BibitemOpen
  \bibinfo {note} {The line $T_{ref}=T$ can be seen as a symmetric line
  analogous to the Nishimori line familiar in spin glass theory \cite {KZ11}.}%
  \bibAnnoteFile{Stop}{Note2}%
\bibitem{Note3}%
  \BibitemOpen
  \bibinfo {note} {Mean-field theory is based on models with quenched disorder
  of the family of the spherical p-spin model. There the so called ``annealed
  approximation'' where the average partition function rather that the average
  free-energy is evaluated turn out to be exact and allows to compute exactly
  the potential.}%
  \bibAnnoteFile{Stop}{Note3}%
\bibitem{Sho12}%
  \BibitemOpen
  \bibfield{author}{%
  \bibinfo {author} {\bibfnamefont{S.}~\bibnamefont{Yaida}},\ }%
  \bibfield{journal}{%
  \bibinfo {journal} {arXiv preprint arXiv:1212.0857}}%
   (\bibinfo {year} {2012})%
  \bibAnnoteFile{NoStop}{Sho12}%
\bibitem{DJS13}%
  \BibitemOpen
  \bibfield{author}{%
  \bibinfo {author} {\bibfnamefont{E.}~\bibnamefont{Dyer}}, \bibinfo {author}
  {\bibfnamefont{J.}~\bibnamefont{Lee}},\ and\ \bibinfo {author}
  {\bibfnamefont{S.}~\bibnamefont{Yaida}},\ }%
  \bibfield{journal}{%
  \bibinfo {journal} {arXiv preprint arXiv:1302.2917}}%
   (\bibinfo {year} {2013})%
  \bibAnnoteFile{NoStop}{DJS13}%
\bibitem{AT78}%
  \BibitemOpen
  \bibfield{author}{%
  \bibinfo {author} {\bibfnamefont{J.}~\bibnamefont{De~Almeida}}\ and\ \bibinfo
  {author} {\bibfnamefont{D.}~\bibnamefont{Thouless}},\ }%
  \bibfield{journal}{%
  \bibinfo {journal} {Journal of Physics A: Mathematical and General}\ }%
  \textbf{\bibinfo {volume} {11}},\ \bibinfo {pages} {983} (\bibinfo {year}
  {1978})%
  \bibAnnoteFile{NoStop}{AT78}%
\bibitem{MPV}%
  \BibitemOpen
  \bibfield{author}{%
  \bibinfo {author} {\bibfnamefont{M.}~\bibnamefont{M{\'e}zard}}, \bibinfo
  {author} {\bibfnamefont{G.}~\bibnamefont{Parisi}},\ and\ \bibinfo {author}
  {\bibfnamefont{M.~A.}\ \bibnamefont{Virasoro}},\ }%
  \emph{\bibinfo {title} {Spin glass theory and beyond}},\ Vol.~\bibinfo
  {volume} {9}\ (\bibinfo {publisher} {World scientific Singapore},\ \bibinfo
  {year} {1987})%
  \bibAnnoteFile{NoStop}{MPV}%
\bibitem{Ca83}%
  \BibitemOpen
  \bibfield{author}{%
  \bibinfo {author} {\bibfnamefont{J.}~\bibnamefont{Cardy}},\ }%
  \bibfield{journal}{%
  \bibinfo {journal} {Physics Letters B}\ }%
  \textbf{\bibinfo {volume} {125}},\ \bibinfo {pages} {470} (\bibinfo {year}
  {1983})%
  \bibAnnoteFile{NoStop}{Ca83}%
\bibitem{CGG99}%
  \BibitemOpen
  \bibfield{author}{%
  \bibinfo {author} {\bibfnamefont{A.}~\bibnamefont{Cavagna}}, \bibinfo
  {author} {\bibfnamefont{J.~P.}\ \bibnamefont{Garrahan}},\ and\ \bibinfo
  {author} {\bibfnamefont{I.}~\bibnamefont{Giardina}},\ }%
  \bibfield{journal}{%
  \bibinfo {journal} {Journal of Physics A: Mathematical and General}\ }%
  \textbf{\bibinfo {volume} {32}},\ \bibinfo {pages} {711} (\bibinfo {year}
  {1999})%
  \bibAnnoteFile{NoStop}{CGG99}%
\bibitem{CGPM03}%
  \BibitemOpen
  \bibfield{author}{%
  \bibinfo {author} {\bibfnamefont{A.}~\bibnamefont{Cavagna}}, \bibinfo
  {author} {\bibfnamefont{I.}~\bibnamefont{Giardina}}, \bibinfo {author}
  {\bibfnamefont{G.}~\bibnamefont{Parisi}},\ and\ \bibinfo {author}
  {\bibfnamefont{M.}~\bibnamefont{M{\'e}zard}},\ }%
  \bibfield{journal}{%
  \bibinfo {journal} {Journal of Physics A: Mathematical and General}\ }%
  \textbf{\bibinfo {volume} {36}},\ \bibinfo {pages} {1175} (\bibinfo {year}
  {2003})%
  \bibAnnoteFile{NoStop}{CGPM03}%
\bibitem{CGP05}%
  \BibitemOpen
  \bibfield{author}{%
  \bibinfo {author} {\bibfnamefont{A.}~\bibnamefont{Cavagna}}, \bibinfo
  {author} {\bibfnamefont{I.}~\bibnamefont{Giardina}},\ and\ \bibinfo {author}
  {\bibfnamefont{G.}~\bibnamefont{Parisi}},\ }%
  \bibfield{journal}{%
  \bibinfo {journal} {Physical Review B}\ }%
  \textbf{\bibinfo {volume} {71}},\ \bibinfo {pages} {024422} (\bibinfo {year}
  {2005})%
  \bibAnnoteFile{NoStop}{CGP05}%
\bibitem{KZ11}%
  \BibitemOpen
  \bibfield{author}{%
  \bibinfo {author} {\bibfnamefont{F.}~\bibnamefont{Krzakala}}\ and\ \bibinfo
  {author} {\bibfnamefont{L.}~\bibnamefont{Zdeborova}},\ }%
  \bibfield{journal}{%
  \Doi{10.1063/1.3506843}{\bibinfo {journal} {The Journal of Chemical
  Physics}}\ }%
  \textbf{\bibinfo {volume} {134}},\ \bibinfo {eid} {034513} (\bibinfo {year}
  {2011}),\ \url{http://link.aip.org/link/?JCP/134/034513/1}%
  \bibAnnoteFile{NoStop}{KZ11}%
\end{thebibliography}%

\end{document}